# Modelling Pre-fatigue, Low-velocity Impact and Fatigue behaviours of Composite Helicopter Tail Structures under Multipoint Coordinated Loading Spectrum

Zheng-Qiang Cheng[a,b], Wei Tan[b], Jun-Jiang Xiong[a,*]

[a] School of Transportation Science and Engineering, Beihang University, Beijing 100191, People's Republic of China (*Corresponding author. E-mail address: jjxiong@buaa.edu.cn)

[b] School of Engineering and Materials Science, Queen Mary University of London, London E1 4NS, United Kingdom

**Abstract:** This paper aims to numerically study the pre-fatigue, low-velocity impact (LVI) and fatigue progressive damage behaviours of a full-scale composite helicopter tail structure under multipoint coordinated loading spectrum. First, a fatigue progressive damage model (PDM) incorporating multiaxial fatigue residual strength degradation rule, fatigue failure criteria based on fatigue residual strength concept and sudden stiffness degradation rule was proposed. Then, an LVI progressive damage model for plain-weave (PW) and unidirectional (UD) composites was developed. Moreover, a full-process analysis algorithm with a reasonable damage transfer strategy for pre-fatigue, LVI and fatigue progressive damage analysis was proposed. Finally, a highly computational efficient and accurate full-scale global-local finite element (FE) model of helicopter tail structure was built to predict strain distribution under two flight working conditions, to predict LVI damage under impact loading, and to assess fatigue damage behaviours under multipoint coordinated loading spectrum. The numerical predictions agree well with test results from this work and literature data, indicating that the developed pre-fatigue, LVI, fatigue PDMs and algorithms, as well as the global-local FE modelling based on shell-to-solid coupling, can effectively analyse the impact damage tolerance of full-scale aircraft structures.

# Nomenclature

| | | | |
|---|---|---|---|
| $C, p, q$ | material parameters in fatigue residual strength degradation model | $R(n)$ | fatigue residual strength pertaining to cyclic stress cycles $n$ |
| $d$ | damage variable | $S_{\max,r}$ | maximum value of nominal cyclic stress at arbitrary stress ratio |
| $E$ | Young's modulus | $S_{\min,r}$ | minimum value of nominal cyclic stress at arbitrary stress ratio |
| $E'$ | damaged Young's modulus | $S_{r_0}$ | maximum absolute value of fatigue stress at specific stress ratio |
| $f$ | intermediate variable | $S_0$ | fatigue endurance limit |
| $G$ | shear modulus | $X$ | static strength for ply |
| $G'$ | damaged shear modulus | $X_0$ | static strength for laminate |
| $G_{ii}^{\mathrm{t}}$ | tensile fracture toughness in $ii$ direction | $\varepsilon^{\mathrm{o}}$ | damage initiation strain |
| $G_{ii}^{\mathrm{c}}$ | compressive fracture toughness in $ii$ direction | $\varepsilon^{\mathrm{f}}$ | damage fracture strain |
| $l^*$ | element characteristic length | $V$ | element volume |
| $n$ | number of cyclic loading cycles | $v$ | Poisson's ratio |
| $r$ | arbitrary stress ratio | $v'$ | damaged Poisson's ratio |
| $r_0$ | specific stress ratio | $\rho$ | density |
| $R$ | fatigue residual strength | $\sigma$ | stress |

**Notation**

| | | | |
|---|---|---|---|
| t | tension | 23 | transverse-through thickness direction |
| c | compression | 1t | longitudinal tension direction |
| 11 | longitudinal direction | 1c | longitudinal compression direction |
| 22 | transverse direction | 2t | transverse tension direction |
| 33 | through-thickness direction | 2c | transverse compression direction |

| 12 | longitudinal-transverse direction | 3t | through-thickness tension direction |
| --- | --- | --- | --- |
| 13 | longitudinal-through thickness direction | 3c | through-thickness compression direction |

# 1 Introduction

Due to the high specific strength and stiffness, fibre-reinforced polymer composites have been widely applied in many parts of civil and military aircrafts, such as fuselages, vertical or horizontal tails and wings[1,2]. Composite aircraft structures used in service are susceptible to low-velocity impact (LVI) damage, especially impact damage from dropping maintenance tools[3]. To meet impact damage tolerance requirements of airworthiness regulations[4], composite aircraft structures with impact damages are requested to prove that these composite structures have sufficient residual strength and life[5]. In recent decades, the majority of research has been done to investigate residual static strength and fatigue strength of composites with LVI damage at a coupon level. It has been reported that the tension after impact (TAI), compression after impact (CAI) and post-impact fatigue (PIF) behaviours and failure mechanism of composite laminates are significantly affected by many factors, such as fibre type[6,7], laminate thickness[8,9], stacking sequence[10-12], impact energy size[13,14] and among others.

The impact damage tolerance behaviours of the composite at large-scale structure level, especially PIF behaviours, are issues that are not completely understood and remain difficult to predict due to the interaction between the geometric nonlinearity of the response, local complex damage modes including fibre breakage, matrix cracking and delamination, and the accumulation of cyclic damage[15]. To address these concerns, Feng et al.[2] experimentally studied the effect of fatigue load on buckling or post-buckling behaviours of stiffened composite panels with impact damage. The results exhibit that the impact damage has no obvious changes during and after fatigue load, but the average failure load of fatigued specimens decrease about 6.1% compared to impacted specimens. In addition, fatigue load has no obvious influence on the buckling load, buckling modes and failure modes. Aoki et al.[16] conducted fatigue tests of hat-shaped stringer stiffened panel with multiple impact damages under mini-TWIST loading spectrum. It has been shown that barely visible impact damage (BVID) and visible impact damage (VID) have not further grown even though under fatigue cycles twice of the design service life, and residual compression strength of the fatigued structure is verified at design limit load level. Tan et al.[17] experimentally investigated the effect of selective stitching on fatigue strengths of stiffened composite panels with visible impact damage on flange and stiffener under constant amplitude fatigue load. The test results indicate that the through-the-thickness stitching is an effective way to improve the post-impact fatigue strength of the stiffened composite structure. Rasuo[18] carried out fatigue tests on the post-impact composite tail rotor blade of a heavy transport multipurpose helicopter. The experimental results show that such a severely damaged blade can perform all its vital functions on the helicopter, even after 65 working hours in extremely difficult flight conditions. Huang and Zhao[19] implemented fatigue tests on composite beams with impact damage under fully reversed torsional fatigue load. It has been shown that both the initial impact damage size and the load history show a significant effect on the damage growth rate and the performance of the post-impact beam. Nevertheless, the above literatures haven't established the link between mesoscopic scale material-level damage mechanisms and the fatigue life of full-scale composite structures after impact.

To shorten the design period and reduce the experimental cost of composite structures, virtual tests performed by finite element (FE) analysis are used to predict the PIF behaviours of composite structures. Rogani et al.[20,21] established a mesoscopic scale FE model where the rod elements are used to model the woven fabric bundles and shell elements for the epoxy matrix. Based on resin matrix and fibre Basquin curves and Miner's damage accumulation rule, the post-impact behaviours of thin carbon/epoxy woven composite laminates and hybrid carbon/epoxy and glass/epoxy woven composite laminates subject to fatigue tensile loading were predicted. The numerical results correlate well with the experimental results in terms of damage propagation scenario, path, and speed. However, this modelling strategy will be computationally expensive in the case of large full-scale composite structures. For the analysis of the PIF behaviours of full-scale composite structures, in the early stages, Attia et al.[22] established a 2D fatigue FE model of the multiple I-stringers stiffened panel with LVI damage. The impact damages including fibre breakage and matrix cracking were modelled by degradation of elastic modulus in the damaged area. The damage growth behaviour of post-impact stringer stiffened panel was then predicted based on the fracture behaviours of composites. Recently, finite-element based progressive damage models (PDMs) have been widely used for investigating fatigue damage mechanisms and predicting the fatigue life of composite structures. Shokrieh and Lessard[23,24] used both residual strength and residual stiffness models to characterise gradual degradation properties of composites, developed fatigue failure criteria based on the Hashin criteria to identify potential failure modes, and adopted sudden stiffness and strength degradation rules to represent composites broken. A fatigue PDM was then established to predict constant amplitude and two-stage High-Low and Low-High fatigue life of composite laminates. Wan et al.[25,26] employed a nonlinear fatigue residual strength model as gradual strength degradation rules, and used fatigue residual strength failure criterion and stiffness reduction material rule in the fatigue PDM. It was then used to assess the fatigue life of composite laminates under spectrum fatigue load and composite helicopter tail structure under multipoint coordinated loading spectrum. Based on the multiaxial fatigue residual strength concept, Xiong et al.[27] developed new 3D Hashin-type fatigue-driven failure criteria, accompanied by sudden stiffness degradation rule, to predict the fatigue life of full-scale composite cabin. Van den Akker et al.[28] adopted cohesive zone based damage model to simulate the adhesive between the skin and stiffener, and used Hashin's damage model to characterise intralaminar damage of stiffened panels neglecting the stiffness and strength decrease under cyclic loads. It was then used to simulate the debonding area growth of co-bonded stiffened panels under constant amplitude fatigue load. Fatigue PDMs are various due to the different combinations of gradual strength/stiffness degradation rules, fatigue failure criteria and sudden material damage metric, and there is no universal PDM that can be applied for all cases, especially PIF fatigue progressive damage analysis. Furthermore, cycle jump strategies are developed to reduce the computational effort of numerical simulations under high-cycle fatigue loading. Turon et al.[29] employed a controlled damage variable level to determine the number of cycle increments in the constitutive fatigue damage model, which is used to predict the delamination in composites under fatigue load. Based on the hypothesis that the stress redistribution due to the fatigue-induced gradual degradation of the material properties can be neglected before sudden damage onset, Russo et al.[30] introduced a smart cycle strategy in the numerical model to investigate the fatigue damage evolution of composite under tension–tension fatigue loading conditions. Moreover, Russo et al.[31] used the increment in the delaminated area to calculate the number of cycle increments in the FaTigue-SMXB approach. It was then used to simulate the compression-compression fatigue behaviour of a stiffened composite panel with an artificial skin-stringer debonding. In addition, the different physical progress between pre-fatigue, LVI and fatigue leads to the increase in the difficulty of simulating the PIF behaviours of composite structures. Therefore, it is necessary to develop an effective fatigue PDM and full-process analysis algorithm for fatigue damage growth and life prediction of composite structure with LVI damage under spectrum fatigue load. In particular, to obtain a high-fidelity FE model for a large composite structure, it is necessary to establish a full-scale FE model, which is costly and time-consuming. Global-local modelling techniques are proposed to establish a global-local FE model with both computational efficiency and accuracy. In the past, global-local modelling approaches including submodelling and shell-to-solid coupling techniques are widely used in the FE model for full-scale composite structures under static load[32-37]. Both techniques are based on a mesh refinement strategy and the detailed introduction of these two approaches are discussed in Section 3. However, there are very limited applications for fatigue and LVI loading cases.

In summary, a large number of experimental and numerical research have been conducted to investigate the PIF behaviours and life prediction of composites from coupon to component level. However, these studies on composite structures, at a primary structure level, with LVI damage are very limited due to the difficulties in establishing effective fatigue PDM and algorithm, and efficient and accurate full-scale global-local FE model. This paper aims to capture the detailed failure mechanisms of the large full-scale helicopter composite tail structures under impact loading and multipoint coordinated loading fatigue spectrum. The main novel contributions herein are: **(i)** A new framework of fatigue progressive damage model coupling multiaxial fatigue residual strength degradation model, fatigue failure criteria and stiffness degradation rule is established. **(ii)** A novel multi-scale finite element model of helicopter composite tail structure was built by the global-local modelling technique. **(iii)** An original full-process analysis algorithm with a reasonable damage transfer strategy for pre-fatigue, LVI and fatigue progressive damage analysis was proposed to predict the detailed failure mechanisms of the full-scale composite helicopter tail structure under multipoint coordinated loading spectrum. Our model opens an avenue to numerically investigate the impact damage tolerance of large full-scale composite aircraft structures.

## 2 Fatigue and LVI progressive damage models

Strength properties of PW and UD composites in longitudinal, transverse, through-thickness, in-plane shear, and out-plane shear directions could degrade under repeated fatigue cyclic loading. Engineering composite structures often suffer from variable amplitude fatigue loads under different stress ratios, so it is essential and desirable to account for the effect of stress ratio on strength degradation. Therefore, based on the fatigue residual strength model of composites[38,39] and modified Goodman curve[25], a multiaxial fatigue residual strength degradation model considering stress ratio effect for PW and UD composites is obtained as,

$$\begin{cases} R_{it}(n) = X_{it} - \left\{ \left[ X_{it} - R_{it}(n-1) \right]^{q_{it}} + C_{it}^{-1} \left( S_{r_0} - S_{0,it} \right)^{-p_{it}} \right\}^{\frac{1}{q_{it}}} & (i=1,2,3) \\ R_{ic}(n) = X_{ic} - \left\{ \left[ X_{ic} - R_{ic}(n-1) \right]^{q_{ic}} + C_{ic}^{-1} \left( S_{r_0} - S_{0,ic} \right)^{-p_{ic}} \right\}^{\frac{1}{q_{ic}}} & (i=1,2,3) \\ R_{ij}(n) = X_{ij} - \left\{ \left[ X_{ij} - R_{ij}(n-1) \right]^{q_{ij}} + C_{ij}^{-1} \left( S_{ij} - S_{0,ij} \right)^{-p_{ij}} \right\}^{\frac{1}{q_{ij}}} & (i,j=1,2,3, i \neq j) \end{cases} \quad (1)$$

with

$$S_{r_0} = \begin{cases} \dfrac{(1-r) X_{it} S_{it\max,r}}{(1-r_0) X_{it} + (r_0 - r) S_{it\max,r}} & (i=1,2,3), (r_0^2 \leq 1, r^2 \leq 1) \\ \dfrac{(r-1) r_0 X_{ic} |S_{ic\min,r}|}{(r_0 - 1) r X_{ic} - (r_0 - r) |S_{ic\min,r}|} & (i=1,2,3), (r_0^2 > 1, r^2 > 1) \end{cases} \quad (2a)$$

$$S_{ij} = \begin{cases} \dfrac{(1-r) X_{ij} S_{ij\max,r}}{(1-r_0) X_{ij} + (r_0 - r) S_{ij\max,r}} & (i,j=1,2,3, i \neq j), (r_0^2 \leq 1, r^2 \leq 1) \\ \dfrac{(r-1) r_0 X_{ij} |S_{ij\min,r}|}{(r_0 - 1) r X_{ij} - (r_0 - r) |S_{ij\min,r}|} & (i,j=1,2,3, i \neq j), (r_0^2 > 1, r^2 > 1) \end{cases} \quad (2b)$$

where $X_{it}$, $X_{ic}$, $X_{ij}$ are the static tension, compression and shear strengths of ply, respectively; $S_{r_0}$ and $S_{ij}$ are the maximum absolute values of fatigue stress at specific stress ratio in normal and shear directions, respectively; $n$ is the number of fatigue cycle; $R_{it}(n)$, $R_{ic}(n)$, $R_{ij}(n)$ are the tension, compression and shear fatigue residual strength after $n$ number of cycles, respectively; $r$ and $r_0$ are the random and specific stress ratios, respectively; $C_{it}, C_{ic}, C_{ij}, p_{it}, p_{ic}, p_{ij}, q_{it}, q_{ic}, q_{ij}, S_{0,it}, S_{0,ic}$, $S_{0,ij}$ are the multiaxial fatigue residual strength degradation model constants. Generally, the constant amplitude tension-tension and compression-compression uniaxial fatigue tests in longitudinal, transverse, through-thickness, in-plane shear, and out-plane shear directions for a PW lamina and a UD lamina are needed to determine all model constants by using best fitting method[39].

Although the Olmedo failure criteria[40] have been successfully used for identifying failure modes of composites under static load, but not in predicting fatigue failure modes because it neglects gradual strength degradation under fatigue load. For this reason, the material strengths in Olmedo's criteria are replaced by multiaxial fatigue residual strength (that is Eq. (1)) to derive fatigue failure criteria for PW and UD composites as listed in Table 1. If the corresponding fatigue failure modes of PW and UD composites happen, the stiffness properties of failed composites are degraded to nearly zero according to sudden stiffness reduction rule[39]. Noticeably, gradually sudden stiffness degradation method is more accurate, but leading to the complexities and difficulties due to strength and stiffness are variable at each fatigue cycle. Compared to large dispersion of fatigue, the calculation error caused by the sudden stiffness reduction rule is acceptable. The degradation of Young's modulus and shear modulus is following the Eq. (3). For sudden stiffness reduction rule, the elastic constants of composites depend on the damage variable $d$. The damage variable has only two values of 0 and 0.99, indicating no fatigue damage and complete fatigue damage, respectively.

$$\begin{cases} E'_{ii} = (1-d_{ii})E_{ii} & (i=1,2,3) \\ v'_{ij} = \dfrac{E'_{ii}v_{ij}}{E_{ii}} & (i,j=1,2,3, i \neq j) \\ G'_{ij} = (1-d_{ij})G_{ij} & (i,j=1,2,3, i \neq j) \end{cases} \tag{3}$$

with

$$d_{ii} = 1-(1-d_{ii}^{t})(1-d_{ii}^{c}) \quad (i=1,2,3) \tag{4a}$$

$$\begin{cases} d_{12} = \max(d_{11}, d_{22}) \\ d_{13} = d_{11} \\ d_{23} = d_{22} \end{cases} \text{for PW}; \quad \begin{cases} d_{12} = d_{11} \\ d_{13} = d_{11} \\ d_{23} = \max(d_{22}, d_{33}) \end{cases} \text{for UD} \tag{4b}$$

where $E'_{ii}, E_{ii}, G'_{ij}, G_{ij}, v'_{ij}, v_{ij}$ are the damaged and undamaged Young's modulus, shear modulus and Poisson's ratio, respectively. As a result, the fatigue PDM of PW and UD composites including multiaxial fatigue residual strength degradation rule, fatigue failure criteria and sudden stiffness degradation rule is established.

Similarly, the LVI failure criteria of PW and UD composites for both intralaminar and interlaminar failure as listed in Table 1 are used to identify potential failure modes under LVI load. The stiffness properties of composites are gradually degraded depending on the current strain value once the LVI failure criteria are met. Each damage variable $d$ monotonically increases from 0 to 1 during the damage evolution, and the damage variables under different failure modes are as follow,

$$d_{ii}^{t(c)} = \frac{\varepsilon_{ii}^{t(c),f} \left( \varepsilon_{ii}^{t(c)} - \varepsilon_{ii}^{t(c),o} \right)}{\varepsilon_{ii}^{t(c)} \left( \varepsilon_{ii}^{t(c),f} - \varepsilon_{ii}^{t(c),o} \right)} \quad (i = 1, 2, 3) \tag{5}$$

with

$$\begin{cases} \varepsilon_{ii}^{t,o} = \dfrac{X_{it}}{E_{it}}; \; \varepsilon_{ii}^{c,o} = \dfrac{X_{ic}}{E_{ic}} & (i = 1, 2, 3) \\ \varepsilon_{ii}^{t,f} = \dfrac{2G_{ii}^{t}}{X_{it} l^{*}}; \; \varepsilon_{ii}^{c,f} = \dfrac{2G_{ii}^{c}}{X_{ic} l^{*}} & (i = 1, 2, 3) \end{cases} \tag{6}$$

where $\varepsilon_{ii}^{t,o}$, $\varepsilon_{ii}^{c,o}$ are the damage initiation strain under tension and compression, respectively; $\varepsilon_{ii}^{t,f}$, $\varepsilon_{ii}^{c,f}$ are the damage fracture strain under tension and compression, respectively; $X_{it}$, $X_{ic}$ are the static tension and compression strengths for ply, respectively; $G_{ii}^{t}$, $G_{ii}^{c}$ are the tensile and compressive fracture toughness, respectively; $l^{*}$ is the characteristic length of element which is determined by $\sqrt[3]{V}$, where $V$ is the element volume. After that, Eq. (3) is used to reduce the elastic constants of PW and UD composites based on the damage variables in each failure mode, and gradual stiffness degradation of composite is achieved. As a result, LVI progressive damage model including LVI failure criteria and gradual stiffness degradation rule is established.

## 3 Global-local FE modelling

To obtain reliable numerical results at a relatively low computational cost, building a highly efficient and accurate global-local FE model is essential for FE analysis of full-scale composite helicopter tail structures. Shell elements are generally used in the global FE model to model the full-scale composite structures to ensure calculation efficiency, meanwhile, solid elements are used in local FE model to model key parts in detail to ensure numerical accuracy.

Most general submodelling technique is the node-based submodelling, which interpolates a nodal displacement field from global model results onto the local submodel nodes. A global analysis is run firstly, and the nodal displacement results in the vicinity of the local submodel boundary are saved as the driven variables, which is used to drive the solution in the local submodel analysis[36,41]. The global model and local submodel analyses are solved separately and sequentially. It is difficult to continuously transfer displacement boundary conditions between global model and local submodel during the solution process for the dynamic loading cases such as fatigue and LVI.

Shell-to-solid coupling modelling as the second global-local modelling technique overcomes this challenge. It assembles constraints that couple the displacement and rotation of each global model's shell node to the average displacement and rotation of the local model's solid surface in the vicinity of the shell node[37,41]. The global and local FE models are analysed concurrently. Hence, we use shell-to-solid coupling modelling technique to establish the global-local FE model of full-scale composite helicopter tail structure.

Table 1  Fatigue and LVI failure criteria for PW and UD composites.

| Failure modes | Fatigue failure criteria | LVI failure criteria |
|---|---|---|
| Warp fibre tension failure of PW or fibre tension failure of UD | $\left(\dfrac{\sigma_{11}}{X_{1t}-[X_{1t}-R_{1t}(n-1)]f_{1t}}\right)^2+\left(\dfrac{\sigma_{12}}{X_{12}-[X_{12}-R_{12}(n-1)]f_{12}}\right)^2+\left(\dfrac{\sigma_{13}}{X_{13}-[X_{13}-R_{13}(n-1)]f_{13}}\right)^2 \geq 1$ | $\left(\dfrac{\sigma_{11}}{X_{1t}}\right)^2+\left(\dfrac{\sigma_{12}}{X_{12}}\right)^2+\left(\dfrac{\sigma_{13}}{X_{13}}\right)^2 \geq 1$ |
| Warp fibre compression failure of PW or fibre compression failure of UD | $\left(\dfrac{\sigma_{11}}{X_{1c}-[X_{1c}-R_{1c}(n-1)]f_{1c}}\right)^2 \geq 1$ | $\left(\dfrac{\sigma_{11}}{X_{1c}}\right)^2 \geq 1$ |
| Weft fibre tension failure of PW or matrix tension failure of UD | $\left(\dfrac{\sigma_{22}}{X_{2t}-[X_{2t}-R_{2t}(n-1)]f_{2t}}\right)^2+\left(\dfrac{\sigma_{12}}{X_{12}-[X_{12}-R_{12}(n-1)]f_{12}}\right)^2+\left(\dfrac{\sigma_{23}}{X_{23}-[X_{23}-R_{23}(n-1)]f_{23}}\right)^2 \geq 1$ | $\left(\dfrac{\sigma_{22}}{X_{2t}}\right)^2+\left(\dfrac{\sigma_{12}}{X_{12}}\right)^2+\left(\dfrac{\sigma_{23}}{X_{23}}\right)^2 \geq 1$ |
| Weft fibre compression failure of PW | $\left(\dfrac{\sigma_{22}}{X_{2c}-[X_{2c}-R_{2c}(n-1)]f_{2c}}\right)^2 \geq 1$ | $\left(\dfrac{\sigma_{22}}{X_{2c}}\right)^2 \geq 1$ |
| Matrix compression failure of UD | $\left(\dfrac{\sigma_{22}}{X_{2c}-[X_{2c}-R_{2c}(n-1)]f_{2c}}\right)^2+\left(\dfrac{\sigma_{12}}{X_{12}-[X_{12}-R_{12}(n-1)]f_{12}}\right)^2+\left(\dfrac{\sigma_{23}}{X_{23}-[X_{23}-R_{23}(n-1)]f_{23}}\right)^2 \geq 1$ | $\left(\dfrac{\sigma_{22}}{X_{2c}}\right)^2+\left(\dfrac{\sigma_{12}}{X_{12}}\right)^2+\left(\dfrac{\sigma_{23}}{X_{23}}\right)^2 \geq 1$ |
| Tension delamination for PW and UD | $\left(\dfrac{\sigma_{33}}{X_{3t}-[X_{3t}-R_{3t}(n-1)]f_{3t}}\right)^2+\left(\dfrac{\sigma_{23}}{X_{23}-[X_{23}-R_{23}(n-1)]f_{23}}\right)^2+\left(\dfrac{\sigma_{13}}{X_{13}-[X_{13}-R_{13}(n-1)]f_{13}}\right)^2 \geq 1$ | $\left(\dfrac{\sigma_{33}}{X_{3t}}\right)^2+\left(\dfrac{\sigma_{23}}{X_{23}}\right)^2+\left(\dfrac{\sigma_{13}}{X_{13}}\right)^2 \geq 1$ |
| Compression delamination for PW and UD | $\left(\dfrac{\sigma_{33}}{X_{3c}-[X_{3c}-R_{3c}(n-1)]f_{3c}}\right)^2+\left(\dfrac{\sigma_{23}}{X_{23}-[X_{23}-R_{23}(n-1)]f_{23}}\right)^2+\left(\dfrac{\sigma_{13}}{X_{13}-[X_{13}-R_{13}(n-1)]f_{13}}\right)^2 \geq 1$ | $\left(\dfrac{\sigma_{33}}{X_{3c}}\right)^2+\left(\dfrac{\sigma_{23}}{X_{23}}\right)^2+\left(\dfrac{\sigma_{13}}{X_{13}}\right)^2 \geq 1$ |

Note: $f_{it}=\left[\dfrac{[X_{it}-R_{it}(n-1)]^{-q_{it}}}{C_{it}(S_{it\max}-S_{0,it})^{p_{it}}}+1\right]^{\frac{1}{q_{it}}}$, $f_{ic}=\left[\dfrac{[X_{ic}-R_{ic}(n-1)]^{-q_{ic}}}{C_{ic}(|S_{ic\min}|-S_{0,ic})^{p_{ic}}}+1\right]^{\frac{1}{q_{ic}}}$ $(i=1,2,3)$; $f_{ij}=\left[\dfrac{[X_{ij}-R_{ij}(n-1)]^{-q_{ij}}}{C_{ij}(S_{ij}-S_{0,ij})^{p_{ij}}}+1\right]^{\frac{1}{q_{ij}}}$ $(i,j=1,2,3, i \neq j)$

**3.1 Structures and materials**

The overall helicopter tail structure is composed of fuselage transition segment, inclined beam, horizontal tail, and tail segment, as shown in Fig. 1. Fuselage transition segment (see Fig. 2(a)) contains 7 frames and 3 surrounding thin walls (top, left, and right), all of which are metallic structures. Inclined beam (see Fig. 2(b)) is composed of 7 frames, 4 surrounding thin walls (top, bottom, left, and right) and an end side wall. Frames C2-C5 are composite frames, while frames C1, C6 and C7 are metallic frames. Both left-side and right-side walls are 4 stringers stiffened skin. For the bottom walls, the first and third bottom walls are honeycomb sandwich composite panels while the second bottom wall is made of metal frame and composite skin. Horizontal tail (see Fig. 2(c)) is made of 9 ribs, 8 stringers, spar, end stringer, and skin, all of which are metallic structures.

Tail segment (see Fig. 3) is composed of 13 frames, 4 surrounding composite stiffened panels (top, bottom, left and right) and middle-side and end-side walls. Frames T2 to T9 are composite frames, while T1, T10 to T13 are metallic frames. The top and bottom stiffened panels are 3 stringers stiffened and 10 stringers stiffened skins, respectively. Both left-side and right-side stiffened panels are 7 stringers stiffened skins. The middle-side and end-side walls are honeycomb sandwich composite panels which are between T10 to T13 frames. The mechanical properties of composite, metal and honeycomb core materials in helicopter tail structure are listed in Tables 2 and 3.

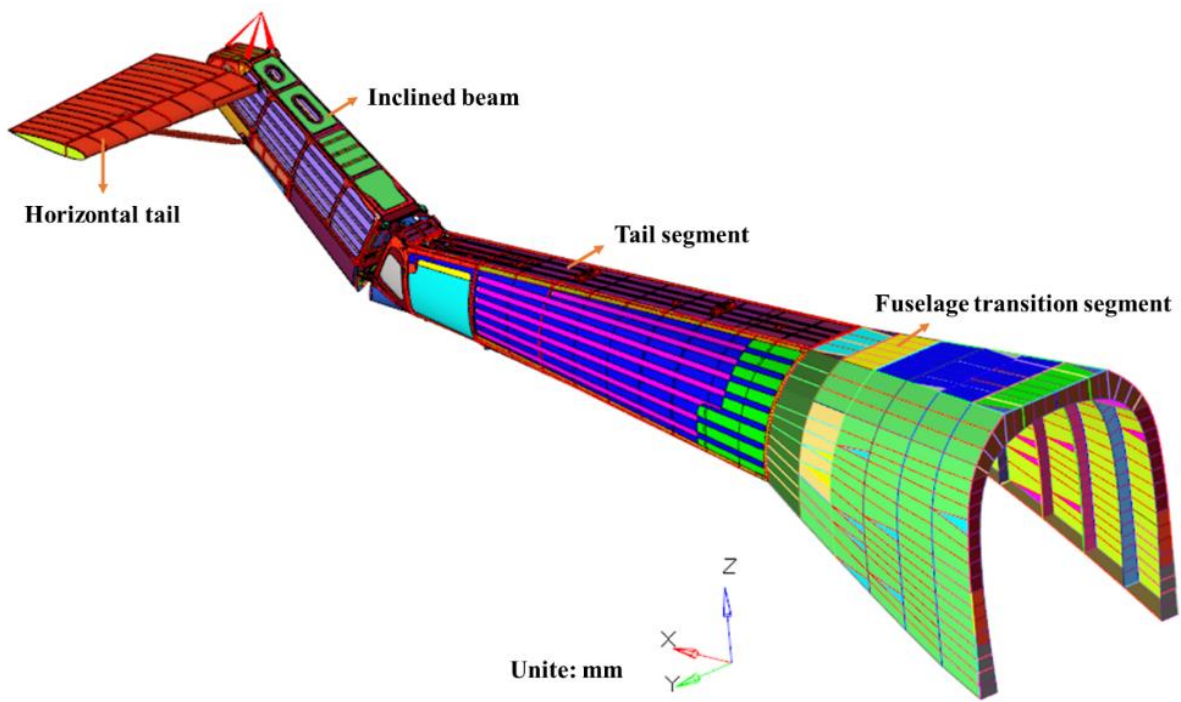

**Fig. 1**　Helicopter tail structure.

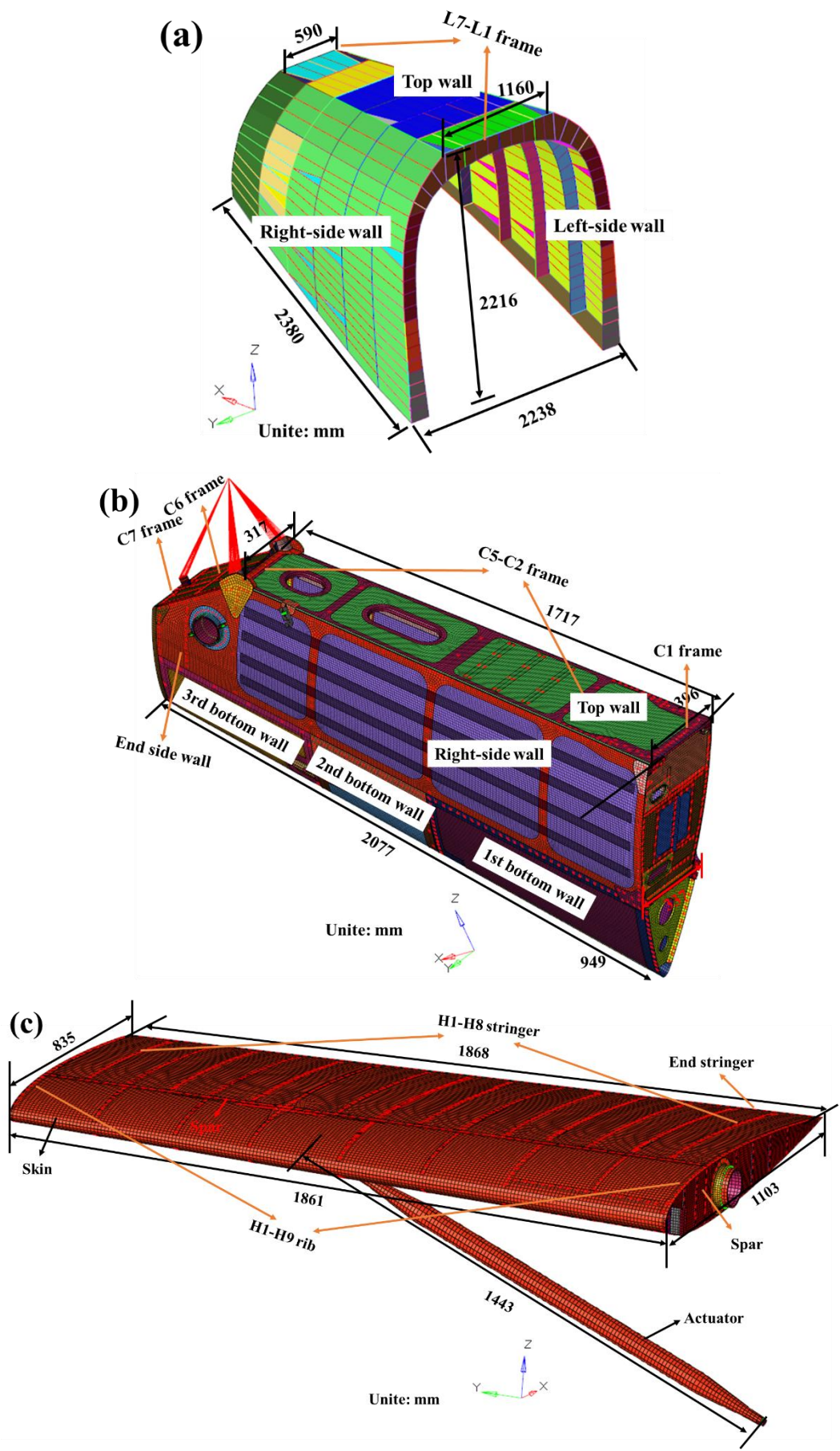

**Fig. 2** Geometry and dimensions: (a) Fuselage transition segment; (b) Inclined beam; (c) Horizontal

tail.

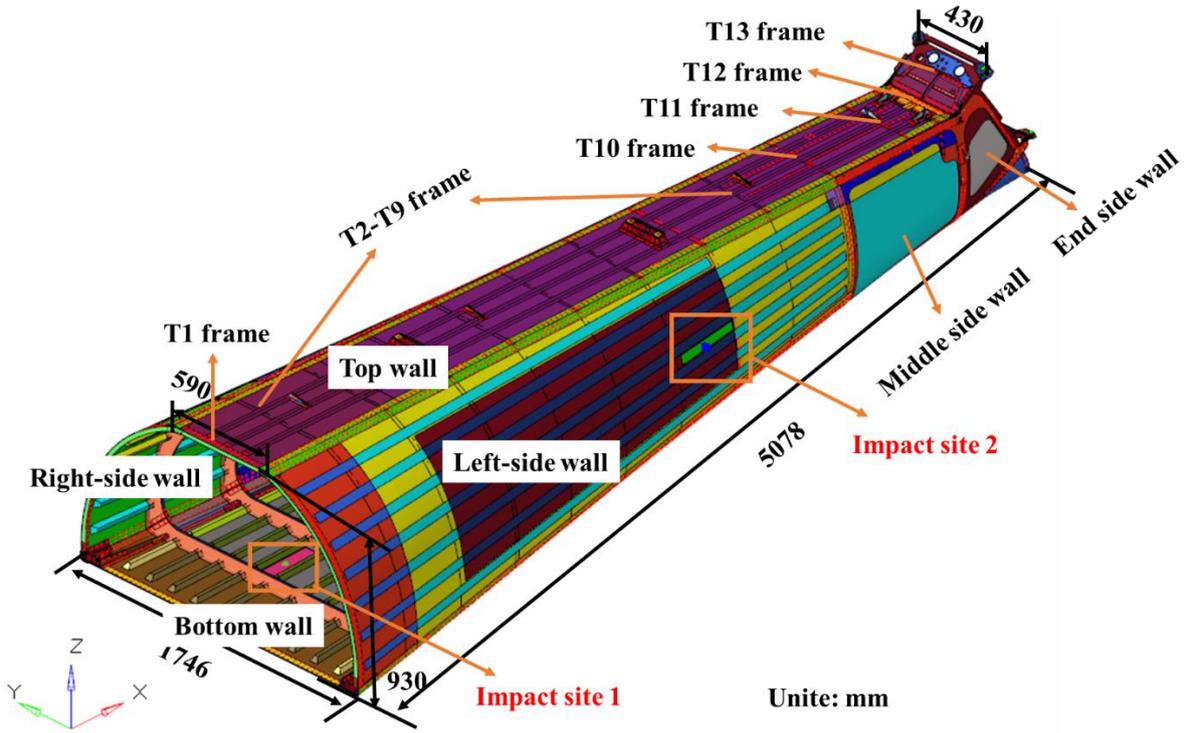

**Fig. 3** Geometry and dimensions of tail segment.

**Table 2** Mechanical properties of PW lamina and UD lamina[10,11,24,42].

| Materials | PW lamina | UD lamina |
|---|---|---|
| Density $\rho$ (g/cm$^3$) | 1.6 | 1.6 |
| Modulus $E$ (GPa) | $E_{11} = E_{22} = 56.97$; $E_{33} = 8.71$<br>$G_{12} = 3.25$; $G_{13} = G_{23} = 2.71$ | $E_{11} = 147$; $E_{22} = E_{33} = 9.0$<br>$G_{12} = G_{13} = 5.0$; $G_{23} = 3.0$ |
| Poisson's ratio | $\nu_{12} = 0.063$; $\nu_{13} = \nu_{23} = 0.3$ | $\nu_{12} = \nu_{13} = 0.3$; $\nu_{23} = 0.42$ |
| Strength (MPa) | $X_{1t} = X_{2t} = 691.62$<br>$X_{1c} = X_{2c} = 557.19$<br>$X_{3t} = 53$; $X_{3c} = 204$<br>$X_{12} = 110.89$; $X_{13} = X_{23} = 66.3$ | $X_{1t} = 2004$; $X_{1c} = 1197$<br>$X_{2t} = X_{3t} = 53$<br>$X_{2c} = X_{3c} = 204$<br>$X_{12} = X_{13} = 137$; $X_{23} = 42$ |
| Fracture energies (kJ/m$^2$) | $G_{11}^t = G_{22}^t = 201$; $G_{33}^t = 0.32$<br>$G_{11}^c = G_{22}^c = 92$; $G_{33}^c = 2.01$ | $G_{11}^t = 92$; $G_{22}^t = G_{33}^t = 0.32$<br>$G_{11}^c = 80$; $G_{22}^c = G_{33}^c = 2.01$ |

Table 3  Mechanical properties of metallic materials and honeycomb core[43,44].

| Materials | Modulus $E$ (MPa) | Poisson's ratio $v$ | Density $\rho$ (g/cm$^3$) |
|---|---|---|---|
| 7050 Aluminium-alloy | 71000 | 0.33 | 2.7 |
| 2618 Aluminium-alloy | 74000 | 0.33 | 2.7 |
| Ti-6Al-4V Titanium-alloy | 110000 | 0.31 | 4.54 |
| D6AC Low-alloy steel | 200000 | 0.32 | 7.85 |
| Nomex | $E_{11}=0.0746$, $E_{22}=0.0746$<br>$E_{33}=121.86$, $G_{12}=0.00329$<br>$G_{23}=13.066$, $G_{13}=20.688$ | 0.9995 | 0.032 |

**3.2 Global-local FE modelling using shell-to-solid coupling method**

According to the geometry and dimensions of helicopter tail structure shown in Figs. 1-3, a full-scale global FE model of helicopter tail structure is established in Hypermesh (Abaqus/Explicit profile). In the FE model of fuselage transition segment (see Fig. 2(a)), three dimensional 2-node truss elements (T3D2) are used to model the stringers, reduced integration 3-node and 4-node shell elements (S3R and S4R) for the skins, and T3D2 and S4R elements for the frames. The elements between stringer and skin share the same nodes at the interface, ensuring a continuous displacement field, so called "co-node" method[26]. In the FE model of inclined beam (see Fig. 2(b)), shell elements S3R and S4R are used to model the skins and frames, and shell elements S3R and S4R and 4-node linear tetrahedron, 6-node linear triangular prism and incompatible mode 8-node linear brick solid elements (C3D4, C3D6, C3D8I) for the stringers. The elements between stringers and skins are also connected by co-node. The rivet connections in the inclined beam are simulated by surface-based coupling constraint and three dimensional 2-node connector element CONN3D2 (see Fig. 4(b)), that is, the nodes of two components that need to be riveted are concentrated to two reference nodes by using the coupling constraint first, the two concentrated reference nodes are then connected through CONN3D2 elements. In the FE model of horizontal tail (see Fig. 2(c)), shell elements S3R and S4R are used to model the skins, stringers, ribs, end stringer, spar, and actuator. The rivet connections in the horizontal tail are also simulated by utilizing the coupling constraint and CONN3D2 elements.

The S3R and S4R elements are used to model the skins and frames, and shell elements (S3R and S4R) and solid elements (C3D4, C3D6, C3D8I) are used to model the stringers in the FE model of tail segment (see Fig. 3). Tie constraint is used for the connections between top skin and stringers, while the co-node is used for connections between left-side skin, right-side skin, bottom skin, and stringers. The top/bottom stiffened panels and the left/right-side stiffened panels are riveted through the lug pieces (see Fig. 4(a) and (b)), that is, the nodes of lug pieces, top/bottom stiffened panels, and left/right-side stiffened panels are concentrated to a specific reference node by using the coupling constraints first, two corresponding reference nodes are then connected through CONN3D2 elements. Frames T2 to T9 are connected to the surrounding stiffened panels by using the tie constraint, while frames T1, T10 to T13 are riveted by employing the coupling constraints and CONN3D2 elements.

The fuselage transition segment and the tail segment are connected by the coupling constraint (see Fig. A1(a)), that is, multiple elements on the tail segment are selected to correspond to one element on the fuselage transition segment, and all nodes of the multiple elements on the tail segment are as the reference node, and then two nodes of one element on the fuselage transition segment are coupled with each reference node by the coupling constraint. The tail segment and the inclined beam are connected by 4 joints which are simulated by using the coupling constraint (see Fig. A1(b)), that is, the nodes inside the lugs at the joints on the tail segment and the nodes at corresponding joints on the inclined beam are coupled to the same reference node. There are 3 connections between the inclined beam and the horizontal tail (see Fig. A1(c)), where the bolt connection is simulated by the coupling constraint, and the rivet connection is simulated by the coupling constraint and CONN3D2 elements.

Based on the service experience of the helicopter, the bottom and left-side skins of the helicopter tail segment are susceptible to low-velocity impact damage, so local FE models in the bottom and left-side skins which involve the LVI events are built simultaneously. A representative local zone is used to determine the local model size. The length of the local model is close to the size of the skin between two adjacent frames, and the width of the local model is the size between two adjacent stringers (see Fig. 3). The local FE model at impact site 1 is shown in Fig. 5, where reduced integration 8-node linear brick solid elements (C3D8R) are used to replace original shell elements of local skin to model a 3D laminated skin. To couple the solid elements of the local FE model with the shell elements of the global FE model, the edge-based surface command is firstly used to create surfaces that contain the shell elements nodes surrounding the local FE model, the element-based surface command is adopted to create surfaces of the solid element nodes at 4 edges of local FE model. The surfaces of the shell element nodes are then coupled with the surfaces of the solid element nodes by the shell-to-solid coupling technique. As a result, a global-local FE model is established. The local FE model at impact site 2 is also built by using the shell-to-solid coupling technique (see Fig. A2). Coarser mesh with element size of 10 mm is used for global FE model. In the local FE model, the element size decreases from 5 mm to 2 mm and 1 mm. The element size of the transition zone is 5 mm and 2 mm, and the element size of the impact zone is 1 mm, which ensures a smooth transition of stress and strain distributions from the shell-to-solid boundary to impact zone. In the local FE models at impact sites 1 and 2, the stacking sequence of laminates are $[(\pm 45)/0_2/(0/90)/0_2/(\pm 45)]$ and $[(\pm 45)/0/(0/90)/0/(\pm 45)]$, respectively, where $(\pm 45)$ and $(0/90)$ plies are the PW layers with a thickness of 0.314 mm, while 0° ply is the UD layer with a thickness of 0.166 mm, and the mechanical properties of PW and UD composites are listed in Table 2.

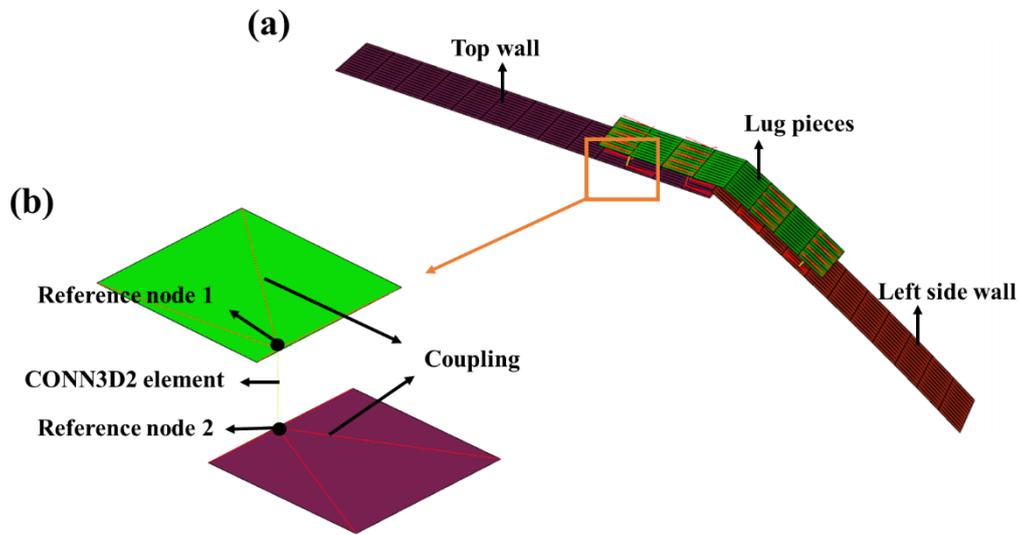

**Fig. 4** Connection technique: (a) Connection between top wall and left side wall; (b) Typical connection technique by coupling constraint and CONN3D2 element.

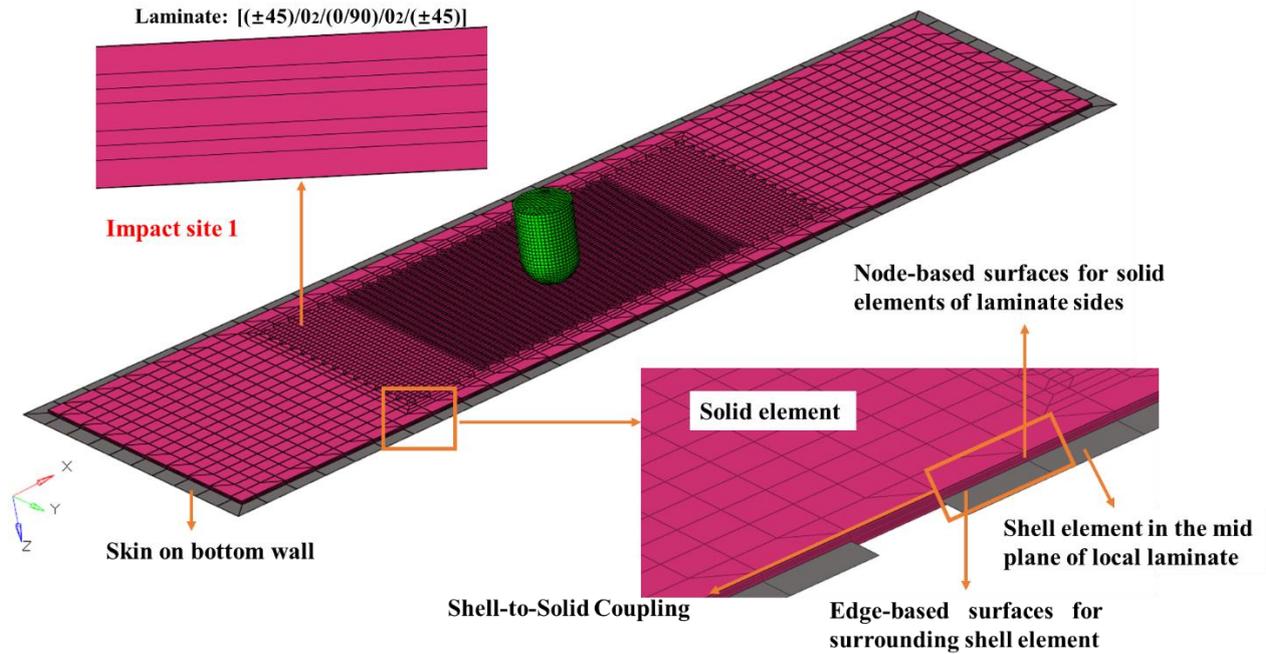

**Fig. 5** Local FE model at impact site 1.

### 3.3 Validation of Global-local FE model

A virtual static test of full-scale helicopter tail structure under multipoint coordinated loading at two typical flight working conditions is analysed, and the predicted strain distribution is compared with the experiments to validate the global-local FE model of full-scale helicopter tail structure. In the physical test, a fixed support constraint is applied on the frame L1 of the fuselage transition segment, and the multipoint coordinated loads of working conditions A and B as shown in Fig. 6 and Table 4 are respectively loaded. The strain distributions on sections S1 to S2 close to the impact point 1 are measured, and the locations and coordinates of the strain measurement points on measuring sections S1 to S2 are shown in Fig. 7. Based on the global-local FE model of full-scale helicopter tail structure (see Figs. 1 and 5), the loading and boundary conditions shown in Fig. 6 and Table 4 are applied, then Abaqus/Standard is used for static analysis, and the details of computational cost are listed in Table 5.

Fig. 8 compares the calculated and experimental strain results in the $x$-direction at two typical working conditions. From Fig. 8, three major findings are as follows:

**i)** Under two flight working conditions, the strain distribution predicted by the global-local FE model agree well with experimental results, indicating the validity and accuracy of the global-local static FE model of full-scale helicopter tail structure.

**ii)** Strain distribution on sections S1 and S2 reaches the minimum value at strain measuring point 3 and the maximum value at strain measuring point 7 or 8 under working condition A, which is mainly related to the applied loading composed of the positive Y-axis loading and negative Z-axis loading.

**iii)** Strain distribution on sections S1 and S2 reaches the minimum value at strain measuring point 3 or 6 and the maximum value at strain measuring point 1 or 8 under working condition B. The top skin is in tension while the bottom skin is in compression, which is caused by positive X-axis loading and negative Z-axis loading without Y-axis loading. In addition, sections S1 and S2 are close to the fuselage transition segment which has an inverted U-shaped boundary rather than a closed boundary condition, thereby limiting the load transfer to the strain measurement points 4 and 5. As a result, the strain distribution of two sections presents a W-shaped line.

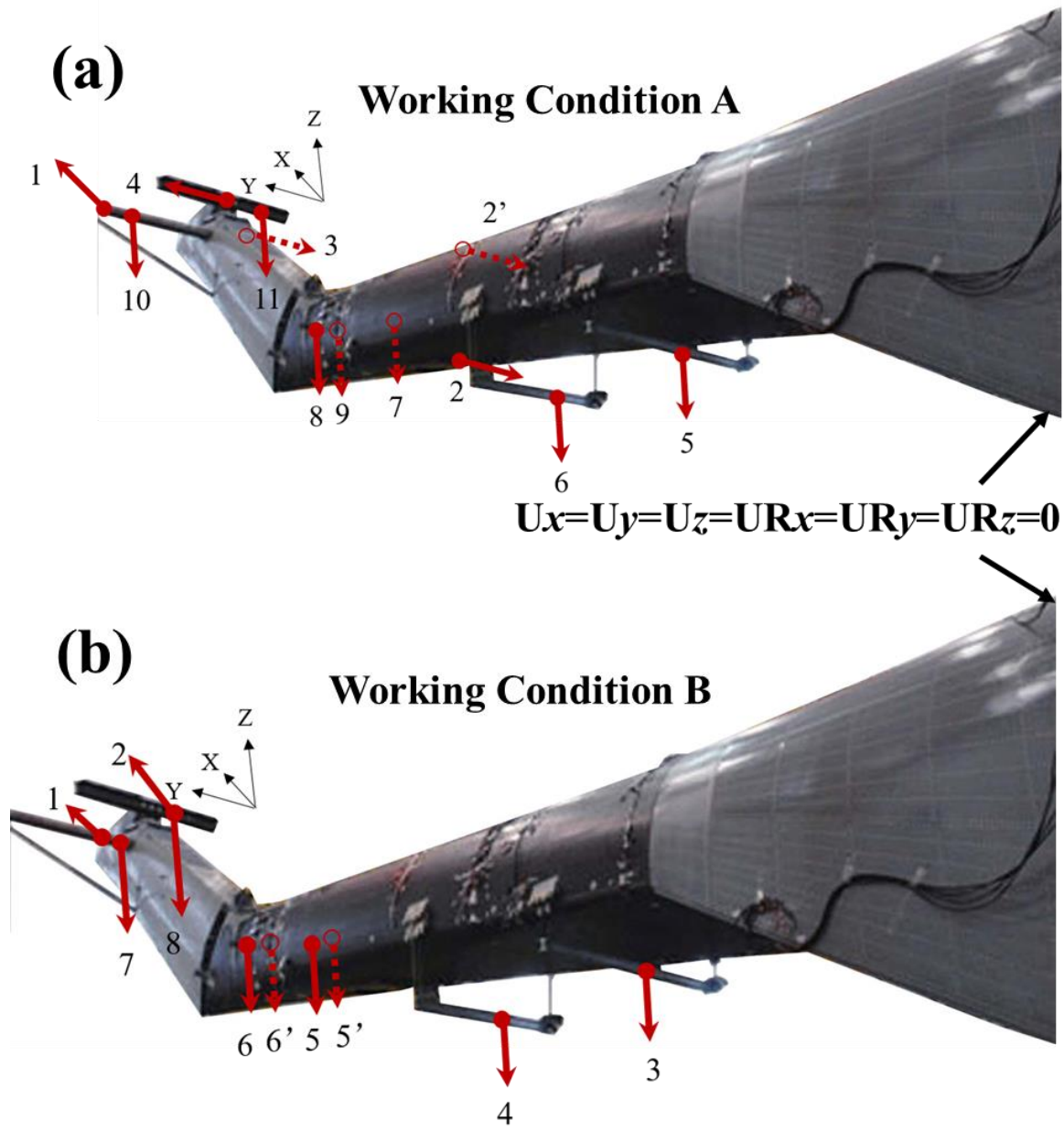

**Fig. 6** Multipoint coordinated loads on full-scale helicopter tail structure: (a) Working condition A; (b) Working condition B.

Table 4  Multipoint coordinated loads under working condition A and B.

| Working condition | Location | Coordinates (mm) | Load (N) | Location | Coordinates (mm) | Load (N) |
|---|---|---|---|---|---|---|
| A | 1 | (18085,1151,3075) | $Fx = 1851.6$ | 6 | (13830,-575.3,1356.3) | $Fz = -778.44$ |
| | 2 | (14650,90,1998) | $Fy = -991.92$ | | (13830,-571.2,1455.3) | $Fz = -778.44$ |
| | | (14650,-240,1233.3) | $Fy = -1852.08$ | | (13830,575.3,1356.3) | $Fz = -684.96$ |
| | 3 | (18085,0,3075) | $Fy = -1351.2$ | | (13830,571.2,1455.3) | $Fz = -684.96$ |
| | 4 | (18085,0,3465) | $Fy = 24543.6$ | 8 | (16267,-255,1804.8) | $Fz = -2570.76$ |
| | 5 | (12600,-758.3,1259.4) | $Fz = -781.56$ | 9 | (16267,255,1804.8) | $Fz = -219.24$ |
| | | (12600,758.3,1259.4) | $Fz = -606.84$ | 10 | (18085,1037.9,3075) | $Fz = -9186$ |
| | 7 | (15470,-331.4,1485.2) | $Fz = -254.4$ | 11 | (18085,-398.9,3075) | $Fz = -6141.6$ |
| B | 1 | (18085,393,3075) | $Fx = 225.6$ | 4 | (13830,-575.3,1356.3) | $Fz = -1140.4$ |
| | 2 | (18085,-399.7,3465) | $Fx = 1008$ | | (13830,-571.2,1455.3) | $Fz = -1140.4$ |
| | 3 | (12600,-758.3,1259.4) | $Fz = -1167.68$ | | (13830,575.3,1356.3) | $Fz = -1140.4$ |
| | | (12600,758.3,1259.4) | $Fz = -851.52$ | | (13830,571.2,1455.3) | $Fz = -1140.4$ |
| | 5 | (15470,-331.4,1485.2) | $Fz = -103.2$ | 6 | (16267,-255,1804.8) | $Fz = -2360.08$ |
| | | (15470,331.4,1485.2) | $Fz = -144.8$ | | (16267,255,1804.8) | $Fz = -2553.52$ |
| | 7 | (18085,319.2,3075) | $Fz = -3596.8$ | 8 | (18085,-399.7,3465) | $Fz = -11347.2$ |

Table 5  Computational details for the static, pre-fatigue, LVI and fatigue.

| Type | Number of elements | Number of nodes | Cores | Memories (GB) | Total CPU time with 48 cores (h) | Physical time (h) |
|---|---|---|---|---|---|---|
| Static | 1318146 | 1211649 | 48 | 48 | 1.50 | 0.03 |
| Pre-fatigue | 1318146 | 1211649 | 48 | 48 | 144.85 | 3.02 |
| LVI | 1323810 | 1214704 | 48 | 48 | 1848.96 | 38.52 |
| Fatigue | 1318146 | 1211649 | 48 | 48 | 243.47 | 5.07 |

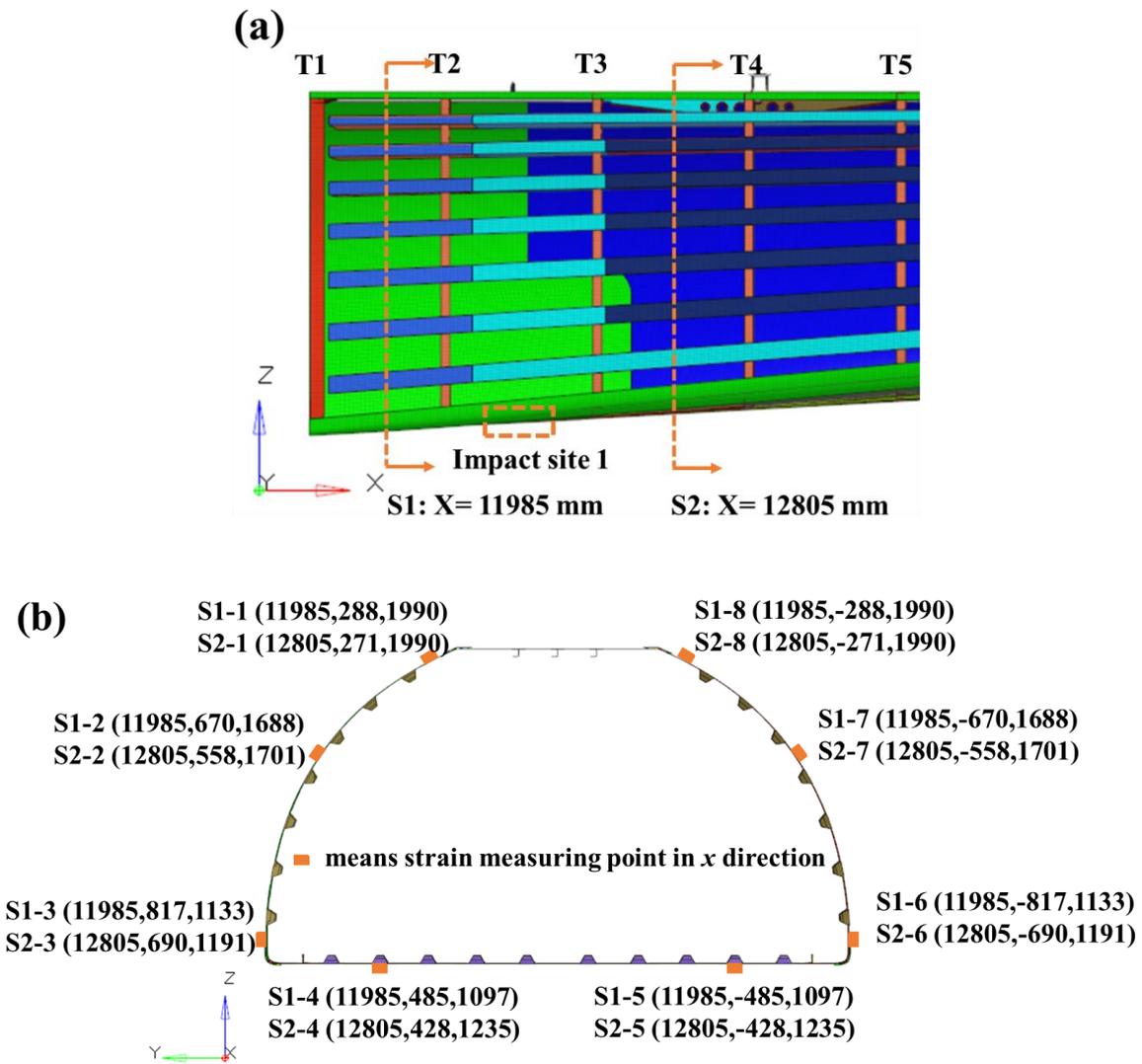

**Fig. 7** Strain measuring sections and coordinates: (a) Strain measuring sections; (b) Strain measuring coordinates.

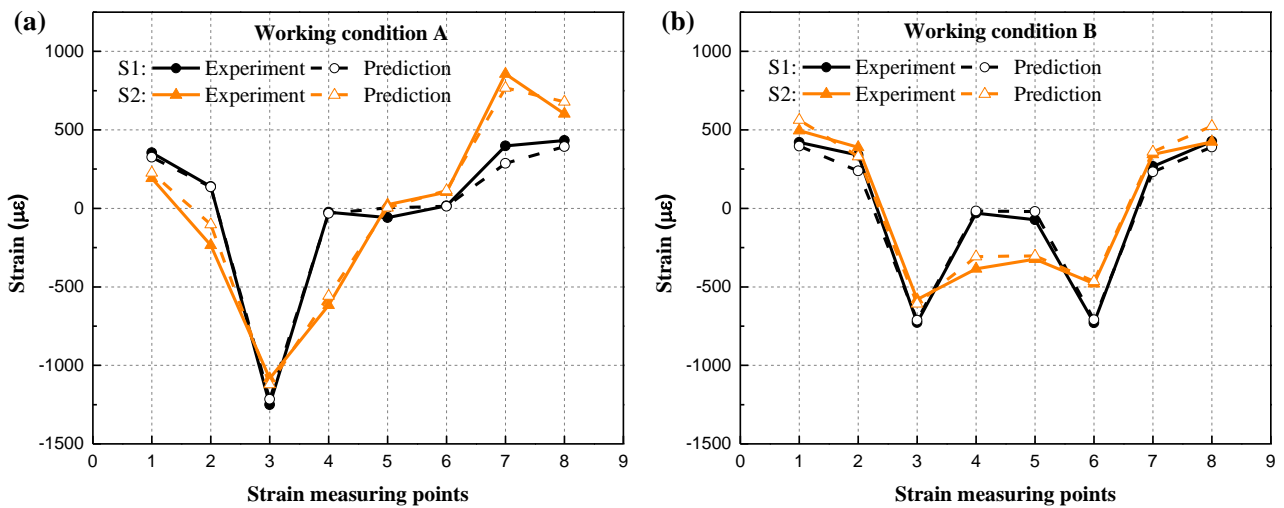

**Fig. 8** Comparison between experimental and predicted strain results in *x*-direction: (a)Working

condition A; (b)Working condition B.

## 4 Progressive damage analysis

### 4.1 Progressive damage analysis algorithm

This paper focuses on investigating the pre-fatigue, LVI and fatigue progressive damage behaviours of full-scale composite helicopter tail structure under multipoint coordinated loading spectrum. Therefore, it is necessary to implement progressive damage analysis on the key parts, that is, the local skins at impact sites, while other non-critical parts do not require progressive damage analysis. The full-process progressive damage analysis of pre-fatigue, LVI and fatigue is shown in Fig. 9. Schematic flowchart for pre-fatigue progressive damage analysis is shown in Fig. 9 (a), and the pre-fatigue progressive damage analysis algorithm is written as a VUMAT subroutine. Firstly, a multipoint coordinated loading spectrum is applied to the global-local FE model of full-scale helicopter tail structure by defining the loading amplitude curve. After stress state analysis of FE model, strength properties of composites are gradually degraded by using multiaxial fatigue residual strength degradation model (see Eq. (1)) and the fatigue failure criteria (see Table 1) are then updated to identify potential fatigue failure. If failure happens, the stiffness of failed element is degraded according to the sudden stiffness reduction rule (see Eq. (3)). The global-local FE stress analysis is re-executed until the fatigue cycle reaches the 48 repeated multipoint coordinated loading spectra. Noticeably, each loading cycle is modelled into the quasi-static loading with the same magnitude as the maximum absolute value of the loading cycle. The cyclic load is applied in two steps: the first step is quasi-static loading to the maximum absolute value of the loading cycle; the second step is to hold the load constant in this cycle, and the analysis pseudo-time increment is assumed to be proportional to the number of loading cycles. The time increment is controlled by mass scaling in Abaqus/explicit to realise the cycle-by-cycle simulation. With such cycle-by-cycle simulation, the pre-fatigue damage is then obtained, including the strength degradation in each material direction and sets of damaged elements.

Passing the damage information from Pre-fatigue stage to the following LVI analysis is very crucial for the whole model. Since the local FE model is only a very small part of the global FE model, the stress distribution on a single lamina in the local FE model is relatively uniform, which is also verified in the pre-fatigue progressive damage analysis. Therefore, we can establish individual degraded material properties for each lamina to model the strength degradation of composites after the pre-fatigue analysis. The various fatigue failure modes saved in state variables are transferred by using the "*Initial Conditions, Type=Solution" command.

According to the flowchart of LVI progressive damage analysis (Fig. 9 (b)), LVI analysis of the global-local FE model is carried out after transferring pre-fatigue damage, and LVI progressive damage algorithm is also written as a VUMAT subroutine. After the stress state of elements is calculated, the failure of elements is identified by LVI failure criteria (see Table 1). If failure happens, the progressive stiffness degradation rule (Eqs. (3)-(5)) is used to gradually degrade the stiffness of the failed element, and the stress state for each element is then updated for further identifying the failure until the end of impact time. Afterwards, we can obtain the LVI damage of the local FE model, including the sets of damage elements in each layer.

The next step is to transfer the pre-fatigue damage and LVI damage to the fatigue PDM. The strength degradation of composites after the pre-fatigue analysis is modelled by applying individual degraded material properties for each layer of laminate. The impact failure modes after LVI analysis saved in state variables are transmitted by using the "*Initial Conditions, Type=Solution command", and the stiffness of the LVI failed elements is suddenly degraded to nearly zero according to the sudden stiffness reduction rule. After that, according to the flowchart of fatigue progressive damage analysis (Fig. 9(c)), a fatigue progressive damage analysis is implemented in the global-local FE model under multipoint coordinated loading spectrum. Similarly, the fatigue progressive damage algorithm is written as a VUMAT subroutine, and its basic principle is consistent with the pre-fatigue progressive damage algorithm.

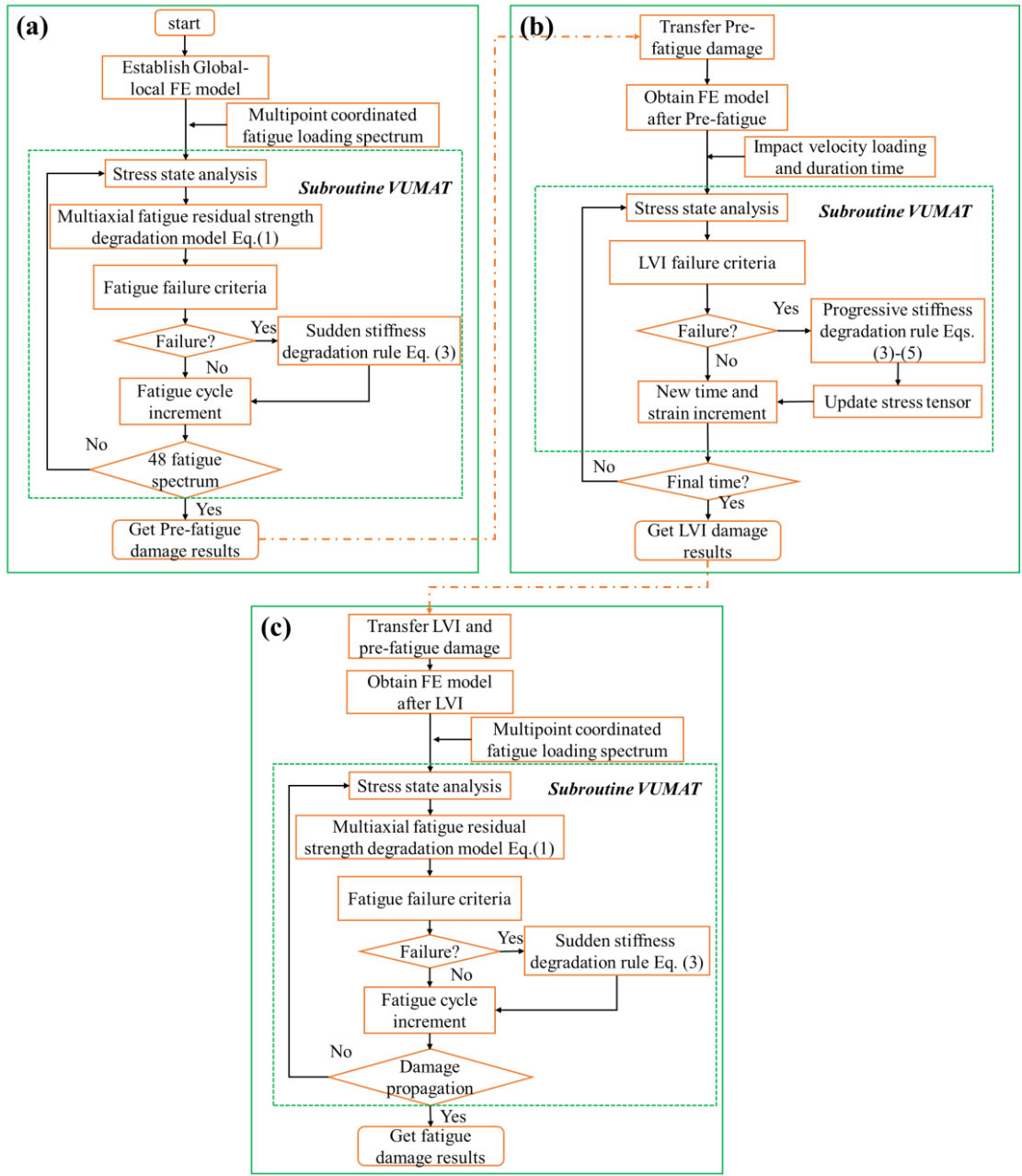

**Fig. 9** Schematic flowcharts of progressive damage analysis: (a) Pre-fatigue analysis; (b) LVI analysis; (c) Fatigue analysis.

**4.2 Pre-fatigue progressive damage analysis**

Multipoint coordinated loading spectrum with reference force of 2000 N is applied to the global-local FE model of full-scale helicopter tail structure (see Figs. 1 and 5) by defining the loading amplitude curve, and a fixed support constraint is applied on the frame L1 of the fuselage transition segment (see Fig. 10). Multipoint coordinated loading fatigue spectrum is determined by equivalently merging the actual load data of helicopter tail structure, and it is equivalent to fatigue loads undertaken by the helicopter tail structure during 1000 flight hours. The fatigue cycles for 1000 flight hours are then simplified and merged into 4000 cycles (shown in Fig. 11) by using the rainflow cycle counting method. According to the pre-fatigue progressive damage analysis algorithm shown in Fig. 9 (a), pre-fatigue analysis is carried out in Abaqus/Explicit. Relevant material properties are listed in Tables 2, 3 and 6, and the details of computational cost are listed in Table 5.

Pre-fatigue progressive damage analysis results are shown in Fig. 12. It can be seen from Fig. 12 that there are no pre-fatigue failed elements at two local laminates after 48 repeated fatigue loading spectra that is equivalent to 48,000 flight hours, which is consistent with the experimental results[26]. It demonstrates that the pre-fatigue progressive damage analysis algorithm and global-local FE model of full-scale helicopter tail structure are efficient and accurate.

Although the 48 repeated fatigue loading spectra do not directly cause the fatigue failure of the local laminate, the strength properties of the internal layers of the laminates decrease irreversibly. The fatigue-driven strength degradation in the local models is shown in Fig. 13, where the explanations of solution-dependent state variables from SDV14 to SDV22 are summarised in Appendix B. From Fig. 13, the following conclusions can be drawn:

**(i)** For Lam 1 and Lam 2, warp fibre tension strength of PW or fibre tension strength of UD, weft fibre tension strength of PW or matrix tension strength of UD, through-thickness tension strength of PW or UD do not degrade (see SDV14, SDV16 and SDV18 in Fig. 13). The reason is that the fatigue limits of PW and UD composites in corresponding material directions are higher than the fatigue stress in the loading spectra (see Table 6), so there is no strength properties reduction.

**(ii)** For warp fibre compression strength of PW or fibre compression strength of UD, in-plane shear strength of PW or UD and out-plane shear strength of PW or UD (see SDV15, SDV20 and SDV22 in Fig. 13), since the fatigue limits for PW plies of 1, 4 and 7 in Lam 1 and 1, 3 and 5 in Lam 2 are higher than the fatigue stress of the loading spectra, no strength degradation happens. However, the fatigue limits for the UD plies of 2, 3, 5, and 6 in Lam 1 and 2 and 4 in Lam2 are lower than the fatigue stress of the loading spectra, so there is strength degradation. For example, the fibre compression strength, in-plane shear strength and out-plane shear strength of ply2 in Lam 1 decreased by 84.47 MPa, 24.97 MPa and 25.92 MPa, respectively. The difference in strength degradation in one layer of the laminate is small due to relatively uniform stress distribution, the maximum strength reduction value is used as the fatigue damage result to be transferred to the LVI analysis.

**(iii)** For weft fibre compression strength of PW or matrix compression strength of UD, through-thickness compression strength, and shear strength in the transverse-through thickness direction (see SDV17, SDV19 and SDV21 in Fig. 13), due to the fatigue limits for PW plies of 1, 4 and 7 in Lam 1 and 1, 3 and 5 in Lam 2 are higher than the fatigue stress in the loading spectra, so there is no strength degradation. Although the fatigue limits for the UD plies of 2, 3, 5, and 6 in Lam 1 and 2 and 4 in Lam2 are lower than the fatigue stress of the loading spectra, the strength degradation is extremely low and can be ignored due to very low fatigue stress in the corresponding stress directions.

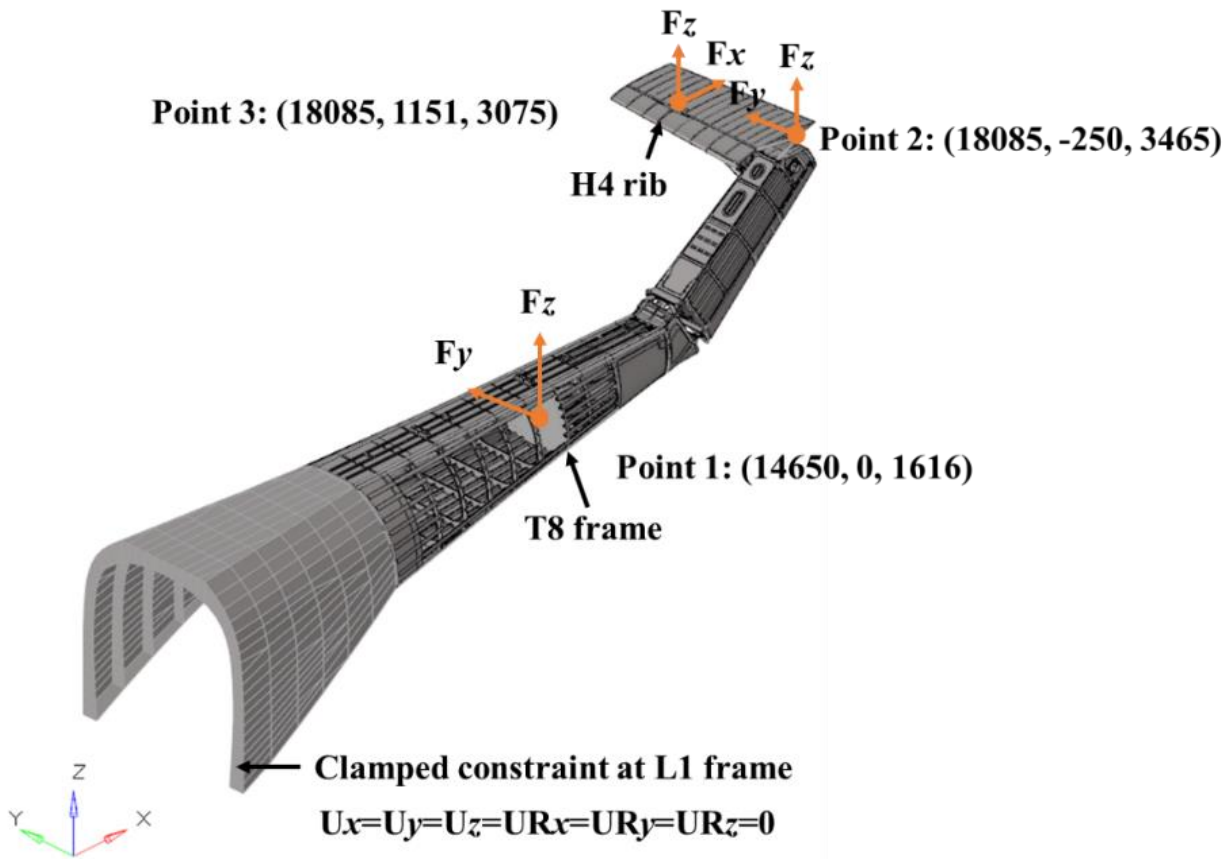

**Fig. 10** Loading points and boundary conditions.

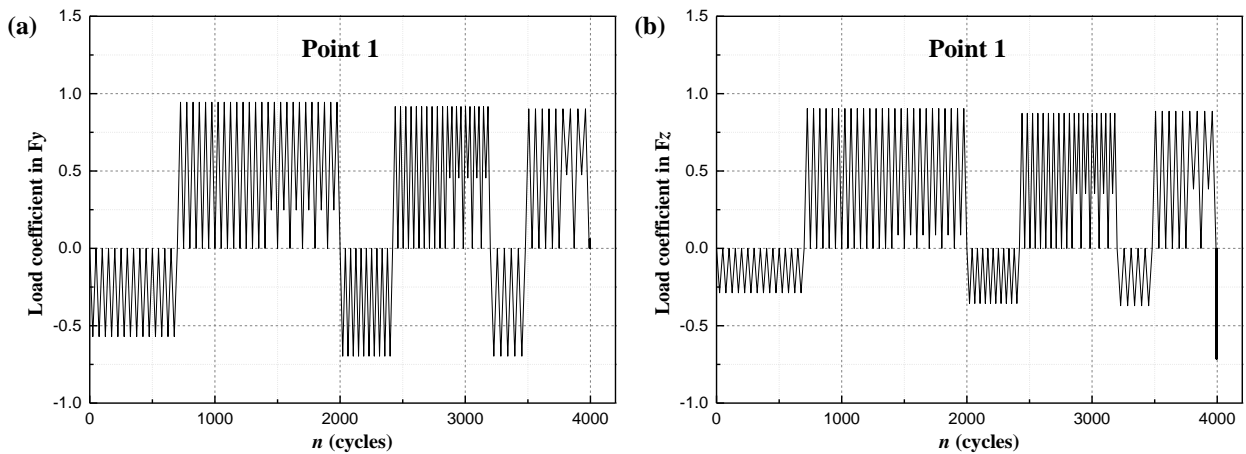

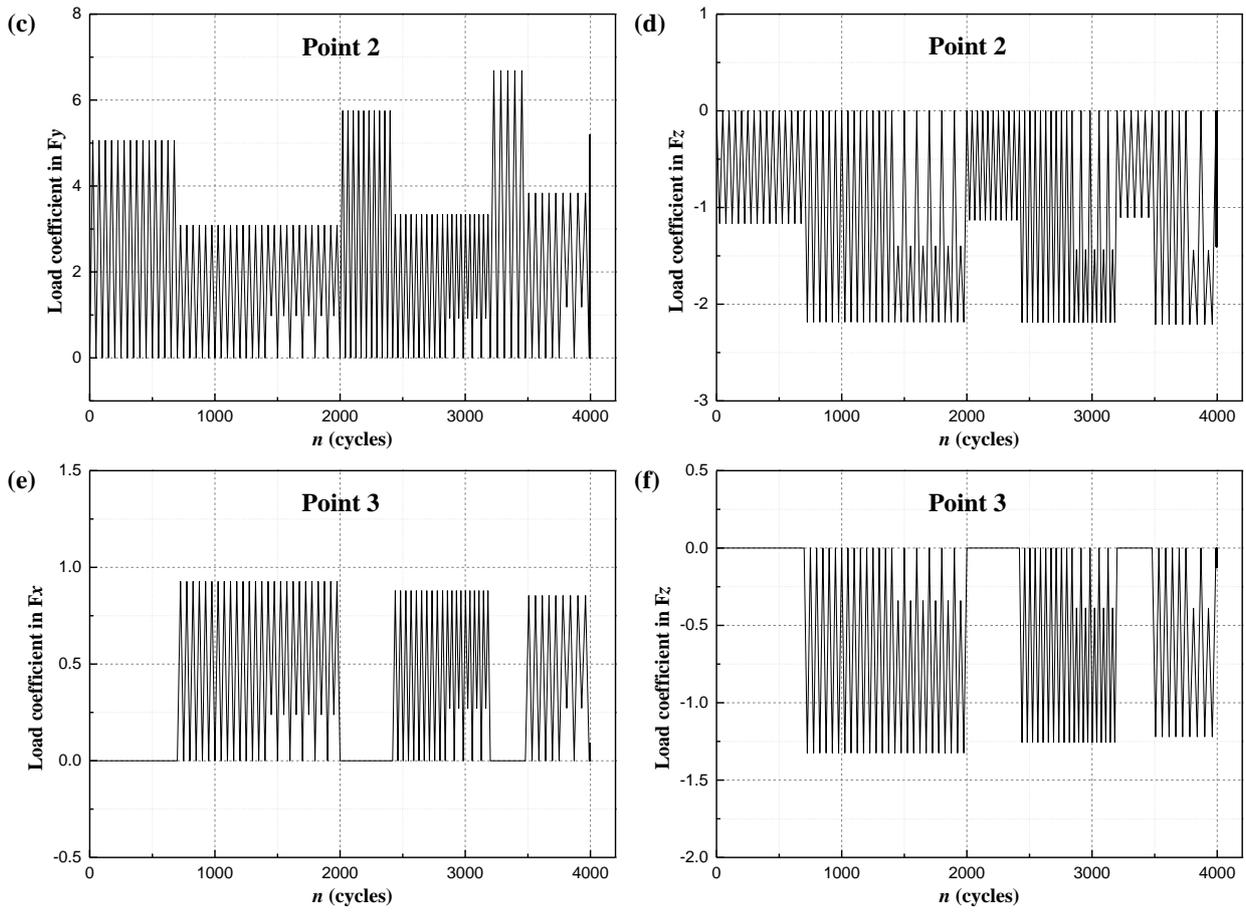

**Fig. 11** Multipoint coordinated loading spectra with reference force of 2000 N: (a) Load coefficient in F$y$ at Point 1, (b) Load coefficient in F$z$ at Point 1, (c) Load coefficient in F$y$ at Point 2, (d) Load coefficient in F$z$ at Point 2, (e) Load coefficient in F$x$ at Point 3, (f) Load coefficient in F$z$ at Point 3.

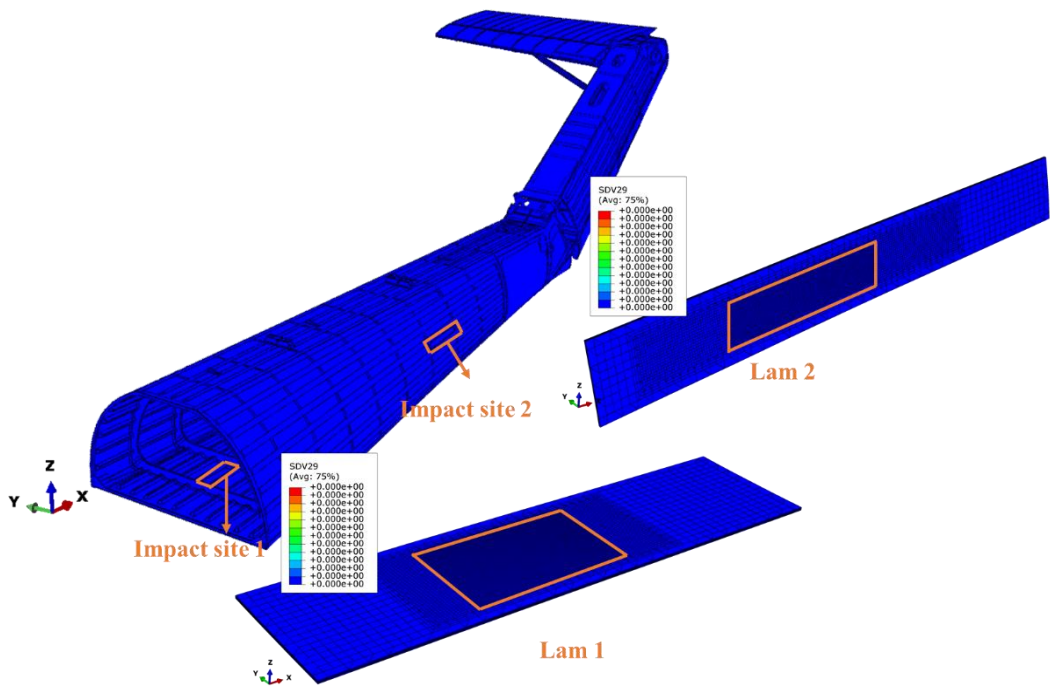

**Fig. 12** Pre-fatigue progressive damage analysis results.

Table 6  Parameters of multiaxial fatigue residual strength degradation model for PW lamina and UD lamina.

| PW lamina | | | | | | UD lamina[24] | | | | |
|---|---|---|---|---|---|---|---|---|---|---|
| $r_0$ | $X_{1t}=X_{2t}$ | $C_{1t}=C_{2t}$ | $p_{1t}=p_{2t}$ | $q_{1t}=q_{2t}$ | $S_{0,1t}=S_{0,2t}$ | $r_0$ | $X_{1t}$ | $C_{1t}$ | $p_{1t}$ | $q_{1t}$ |
| 0.05 | 691.62 | $7.76 \times 10^8$ | -3.79 | 1.86 | 501.56 | 0.10 | 2004.00 | $4.73 \times 10^{12}$ | -4.93 | 2.89 |
| $r_0$ | $X_{1c}=X_{2c}$ | $C_{1c}=C_{2c}$ | $p_{1c}=p_{2c}$ | $q_{1c}=q_{2c}$ | $S_{0,1c}=S_{0,2c}$ | $r_0$ | $X_{1c}$ | $C_{1c}$ | $p_{1c}$ | $q_{1c}$ |
| 10 | 557.19 | $8.70 \times 10^9$ | -3.45 | 0.74 | 283.38 | 10 | 1197.00 | $4.57 \times 10^1$ | -0.44 | 2.16 |
| $r_0$ | $X_{3t}$ | $C_{3t}$ | $p_{3t}$ | $q_{3t}$ | $S_{0,3t}$ | $r_0$ | $X_{2t}=X_{3t}$ | $C_{2t}=C_{3t}$ | $p_{2t}=p_{3t}$ | $q_{2t}=q_{3t}$ |
| 0.1 | 53.00 | $6.71 \times 10^{-1}$ | -1.88 | 5.24 | 17.39 | 0.10 | 53.00 | $6.71 \times 10^{-1}$ | -1.88 | 5.24 |
| $r_0$ | $X_{3c}$ | $C_{3c}$ | $p_{3c}$ | $q_{3c}$ | $S_{0,3c}$ | $r_0$ | $X_{2c}=X_{3c}$ | $C_{2c}=C_{3c}$ | $p_{2c}=p_{3c}$ | $q_{2c}=q_{3c}$ |
| 10 | 204.00 | $2.83 \times 10^{34}$ | -16.12 | 2.63 | 0.00 | 10 | 204.00 | $2.83 \times 10^{34}$ | -16.12 | 2.63 |
| $r_0$ | $X_{12}$ | $C_{12}$ | $p_{12}$ | $q_{12}$ | $S_{0,12}$ | $r_0$ | $X_{12}=X_{13}$ | $C_{12}=C_{13}$ | $p_{12}=p_{13}$ | $q_{12}=q_{13}$ |
| 0.05 | 110.89 | $2.28 \times 10^{18}$ | -9.02 | 0.38 | 41.71 | 0.10 | 137.00 | $1.20 \times 10^{-3}$ | -0.97 | 5.89 |
| | | | | | | $r_0$ | $X_{23}$ | $C_{23}$ | $p_{23}$ | $q_{23}$ |
| | | | | | | 0.10 | 42.00 | $7.07 \times 10^8$ | -5.73 | 4.10 |

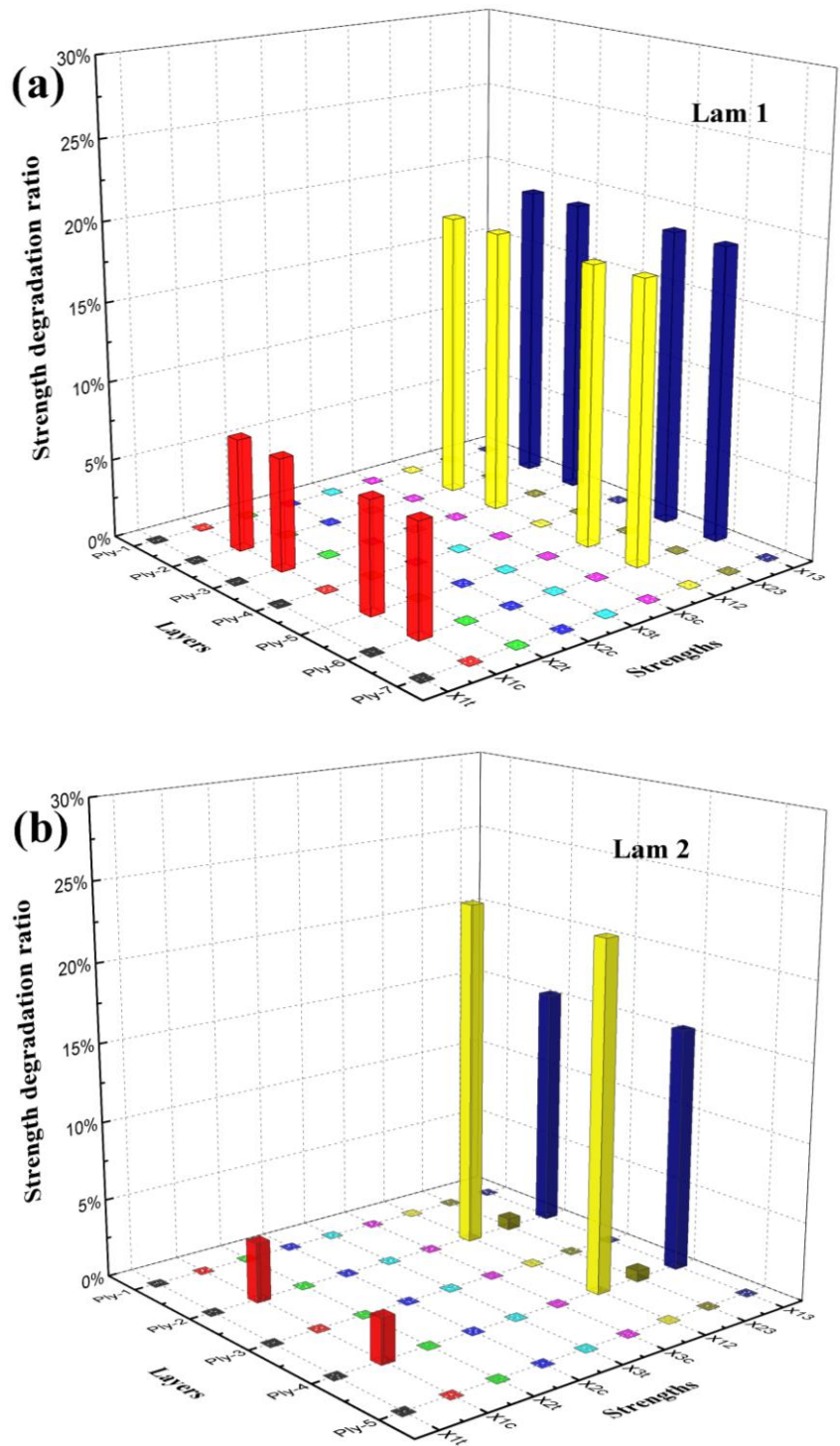

**Fig. 13** Fatigue-driven strength degradation after 48 repeated fatigue loading spectra: (a) Lam 1; (b) Lam 2.

### 4.3 LVI progressive damage analysis

Based on the global-local FE model of full-scale helicopter tail structure, rigid elements R3D3 and R3D4 are used to model the hemispherical impactor with a diameter of 16 mm and mass of 5.61 kg. The impact velocity is loaded at the tip of the impactor for modelling the normal impact energies of 22.2 J and 5.18 J at impact sites 1 and 2 (see Figs. 5 and A2), respectively. The fixed support constraint is also applied on the frame L1 of the fuselage transition segment. The general contact algorithm is used to simulate any impactor-ply or ply-ply contact which may arise in the model and a friction coefficient of 0.3 is used for all potential contact. LVI analysis is implemented in Abaqus/Explicit according to the LVI progressive damage analysis algorithm shown in Fig. 9 (b). Relevant material properties are listed in Tables 2 and 3, and the details of computational cost are listed in Table 5.

LVI total damage results for Lam 1 and Lam 2 are shown in Fig. 14, and various LVI failure modes for Lam 1 are shown in Fig. 15, where the explanations of solution-dependent state variables from SDV1 to SDV6 are summarised in Appendix B. From Figs. 14 and 15, three main findings can be obtained as follows:

**(i)** Predicted LVI damage footprint of Lam 1 and Lam 2 can be deemed as the circle pattern (see Fig. 14), which is consistent with general visual inspection results, and numerical LVI damage area agrees well with the ultrasound C-scan results[26], demonstrating the effectiveness of LVI progressive damage algorithm and global-local LVI FE model of helicopter tail structure. In addition, the predicted overall damage area of Lam 1 is significantly larger than that of Lam 2 because the impact energy for Lam 1 is about 4 times that for Lam 2.

**(ii)** For PW plies of 1, 4 and 7 in Lam 1 (see Fig. 15(a)-15(d)), due to the front first ply and bottom seventh ply are (±45) layers, warp fibre and weft fibre mainly fail in tension mode with the direction of 45°, and there are lager warp fibre and weft tension fibre failure at seventh ply because this layer is subjected to greater tensile stress. The middle fourth ply is a (0/90) layer, and its failure shape presents circular. PW plies of 1, 3 and 5 in Lam 2 have similar above-mentioned failure results.

**(iii)** There is only fibre tension failure but no fibre compression failure (see Fig. 15(a) and 15(b)), and the most severe failure mode is matrix tension failure (see Fig. 15(c) and 15(d)) for four 0° layers of Lam 1 including plies 2, 3, 5, and 6. The bottom sixth ply undergoes greater tensile stress, leading to higher matrix tension failure. Tension and compression delamination mainly occur in the middle layers from ply 2 to ply 6 (see Fig. 15(e) and 15(f)). Similarly, the fibre failures, matrix failures, and delamination in 0° layers of Lam 2 are consistent with the aforementioned failure results.

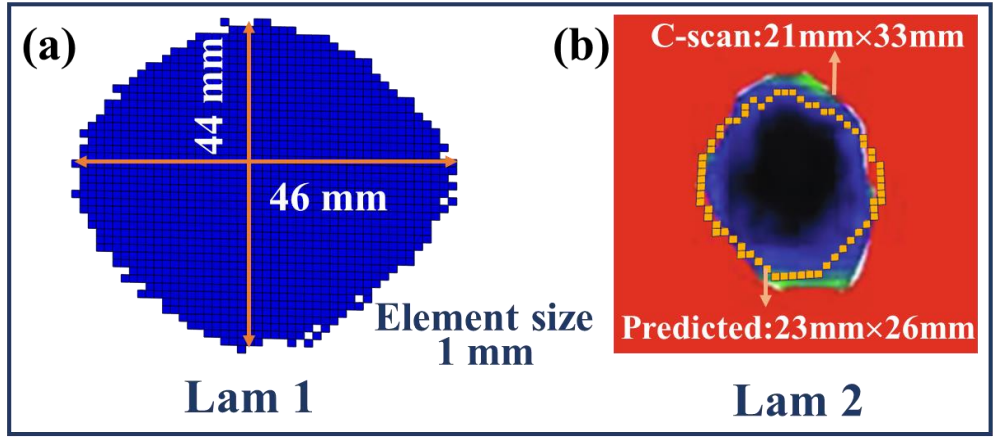

**Fig. 14** LVI damage results for Lam 1 and Lam 2: (a) predicted LVI damage footprint for Lam 1, note that the C-scan damage contour for Lam 1 is not available; (b) predicted and measured LVI damage footprint for Lam 2.

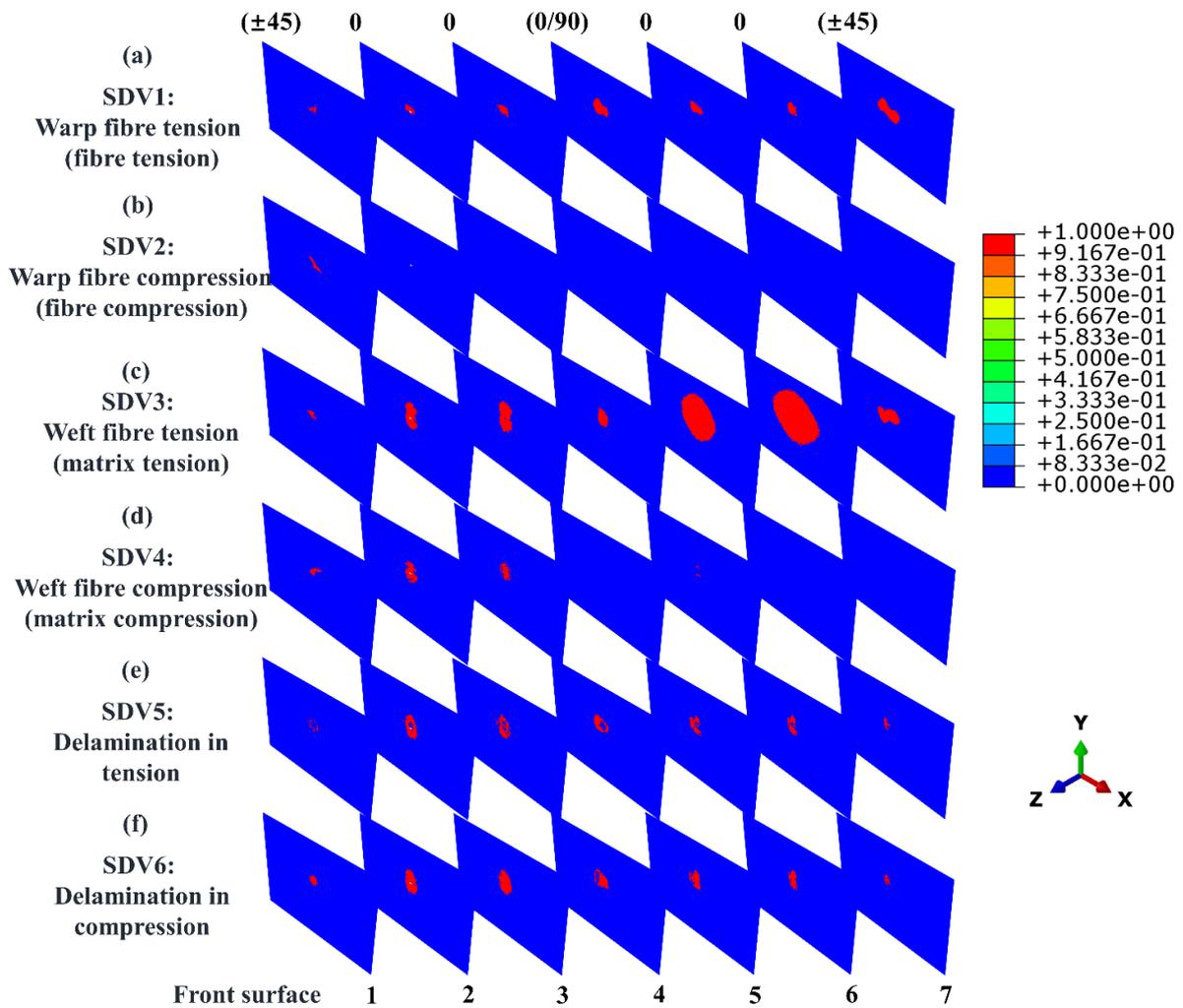

**Fig. 15** LVI failure modes for Lam 1: (a) SDV1: warp fibre tension failure of PW or fibre tension failure of UD; (b) SDV2: warp fibre compression failure of PW or fibre compression failure of UD;

(c) SDV3: weft fibre tension failure of PW or matrix tension failure of UD; (d) SDV4: weft fibre compression failure of PW or matrix compression failure of UD; (e) SDV5: tension delamination for PW or UD; (f) SDV6: compression delamination for PW or UD.

**4.4 Fatigue progressive damage analysis**

The loading and boundary conditions of the global-local fatigue FE model of full-scale helicopter tail structure are consistent with the pre-fatigue FE model of Section 5.2, and the basic principle of fatigue and pre-fatigue progressive damage algorithms are similar. The only difference is that the pre-fatigue damage and LVI damage need to be transferred to the fatigue PDM according to the method described in Section 4.1 before fatigue analysis. Fatigue damage growth behaviours is predicted in Abaqus/Explicit according to the fatigue progressive damage analysis algorithm shown in Fig. 9 (c). Relevant material properties are listed in Tables 2, 3 and 6, and the details of computational cost are listed in Table 5.

Fatigue damage growth behaviours of typical layers in the Lam 1 are shown in Fig. 16 under multipoint coordinated loading spectrum as shown in Fig. 11. From Fig. 16, three main results are summarized as follows:

**(i)** All fatigue failure modes in Lam 1 propagate obviously after 5 repeated fatigue loading spectra. Following the current "no growth" design concept of composite structures, meaning no further damage growth is allowed after the damage is initiated[5], the predicted fatigue life of full-scale helicopter tail structure is 48 + 5 repeated fatigue loading spectra, which is equivalent to 53,000 flight hours. The predicted fatigue life corelate well with experimental fatigue life of 56,000 flight hours[26]. It demonstrates that fatigue progressive damage analysis algorithm and global-local fatigue FE model can effectively predict the fatigue damage behaviour of full-scale helicopter tail structure under multipoint coordinated loading spectrum.

**(ii)** For the PW plies, such as the first ply (see Figs. 16(a)), the warp fibre and weft fibre dominate the fatigue failure. After 5 repeated fatigue loading spectra, warp fibre tension and compression, weft fibre tension and compression, and delamination in tension and compression grow to the same extent damage due to the mixed effect of tensile and compressive loads. There is no more damage growth at 11 repeated fatigue loading spectra.

**(iii)** For the UD plies, such as the fifth ply (see Figs. 16(b)). The initial matrix tension failure is dominant. Due to the mixed effect of tensile and compressive loads, matrix compression failure and delamination grow quickly, leading to strength and stiffness degradation in that region. Fibre compression failure mode (SDV2) is then more likely to grow under the compressive loads of the loading spectrum after 5 repeated fatigue loading spectra. Large area propagation of matrix failure and delamination happens, and tension and compression fibre failure modes (SDV1 and SDV2) also grow obviously after 11 repeated fatigue loading spectra.

From the progressive damage analysis results of pre-fatigue, LVI and fatigue presented from Section 4.2 to Section 4.4, it can prove that the developed pre-fatigue, LVI, fatigue PDMs and algorithms, as well as the global-local FE modelling technique, can effectively analyse impact damage tolerance of full-scale composite structures, opening a new avenue to implement damage tolerance design at composite structure level in engineering practice.

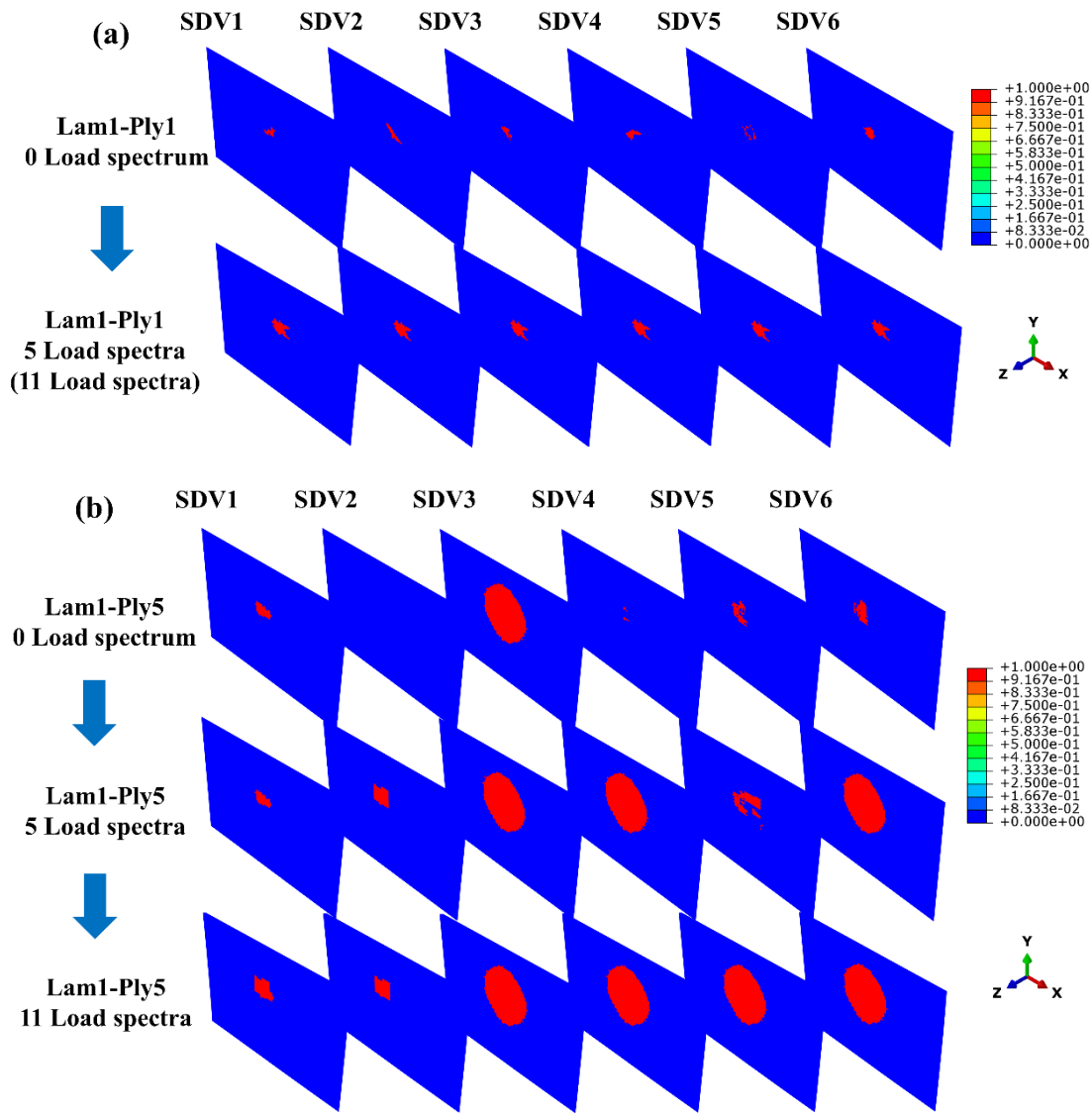

**Fig. 16** Fatigue damage propagation for Lam 1 under multipoint coordinated loading spectrum: (a) Ply1;(b) Ply5.

## 5 Conclusions

This paper numerically studies the progressive damage behaviour of the pre-fatigue, low-velocity impact (LVI), fatigue for full-scale composite helicopter tail structure under multipoint coordinated loading spectrum. The following conclusions can be drawn from this investigation:

**(i)** A highly efficient and accurate global-local finite element (FE) model has been established for full-scale helicopter tail structure. The predicted strain distribution agrees well with experimental results under two typical flight working conditions, demonstrating the validity of the global-local FE model of helicopter composite tail structure based on the shell-to-solid coupling technique.

**(ii)** Fatigue progressive damage model incorporating multiaxial fatigue residual strength degradation rule, fatigue failure criteria and sudden stiffness degradation rule are proposed. LVI progressive damage model including LVI failure criteria and gradual stiffness degradation rule is developed. Moreover, a full-process analysis algorithm with a reasonable damage transfer strategy for pre-fatigue, LVI and fatigue progressive damage analysis was proposed.

**(iii)** Based on the proposed progressive damage models and algorithms and built global-local FE model of the full-scale helicopter tail structure, pre-fatigue damage under multipoint coordinated loading spectrum, LVI damage under impact load, and fatigue damage growth and fatigue life under multipoint coordinated loading spectrum are predicted. The prediction results correlate well with the experiments, indicating that the developed pre-fatigue, LVI, fatigue progressive damage models and algorithms, as well as the global-local FE modelling based on shell-to-solid coupling, can effectively analyse the impact damage tolerance of full-scale aircraft structures.

Various possibilities can be envisaged to continue this investigation:

(i) It seems necessary for more experimental results of pre-fatigue, LVI and PIF composites to further validate the mixed algorithm and global-local FE modelling technique.

(ii) Parametric studies (e.g., stacking sequence, impact energy level, occurrence probability of LVI site and time, size effect of local FE model, etc.) are more helpful and beneficial to deeply understand progressive pre-fatigue, LVI and PIF damage mechanism and to deduce general conclusions.

(iii) It is desirable to investigate the smart cycle jump strategy for further improving computational efficiency in progressive fatigue damage modelling.

## Acknowledgements

This project was supported by the National Natural Science Foundation of China [Grant No. 51875021] and the China Scholarship Council [Grant No. 202006020210]. W. Tan acknowledges financial support from the EPSRC [Grant EP/V049259/1] and the European Commission Graphene Flagship Core Project 3 (GrapheneCore3) under [Grant No. 881603]. The authors also acknowledge the Queen Mary's Apocrita HPC facility. http://doi.org/10.5281/zenodo.438045.

For the purpose of open access, the author has applied a creative commons attribution (CC BY) licence (where permitted by UKRI, 'open government licence' or 'creative commons attribution no-derivatives (CC BY-ND) licence' may be stated instead) to any author accepted manuscript version arising.

## Appendix A

Additional connection strategies are shown in Fig. A1, and local finite element (FE) model at impact site 2 is shown in A2.

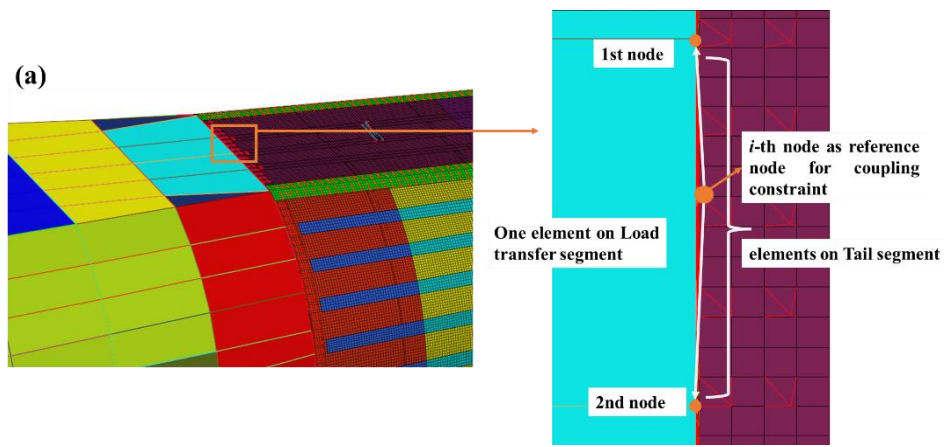

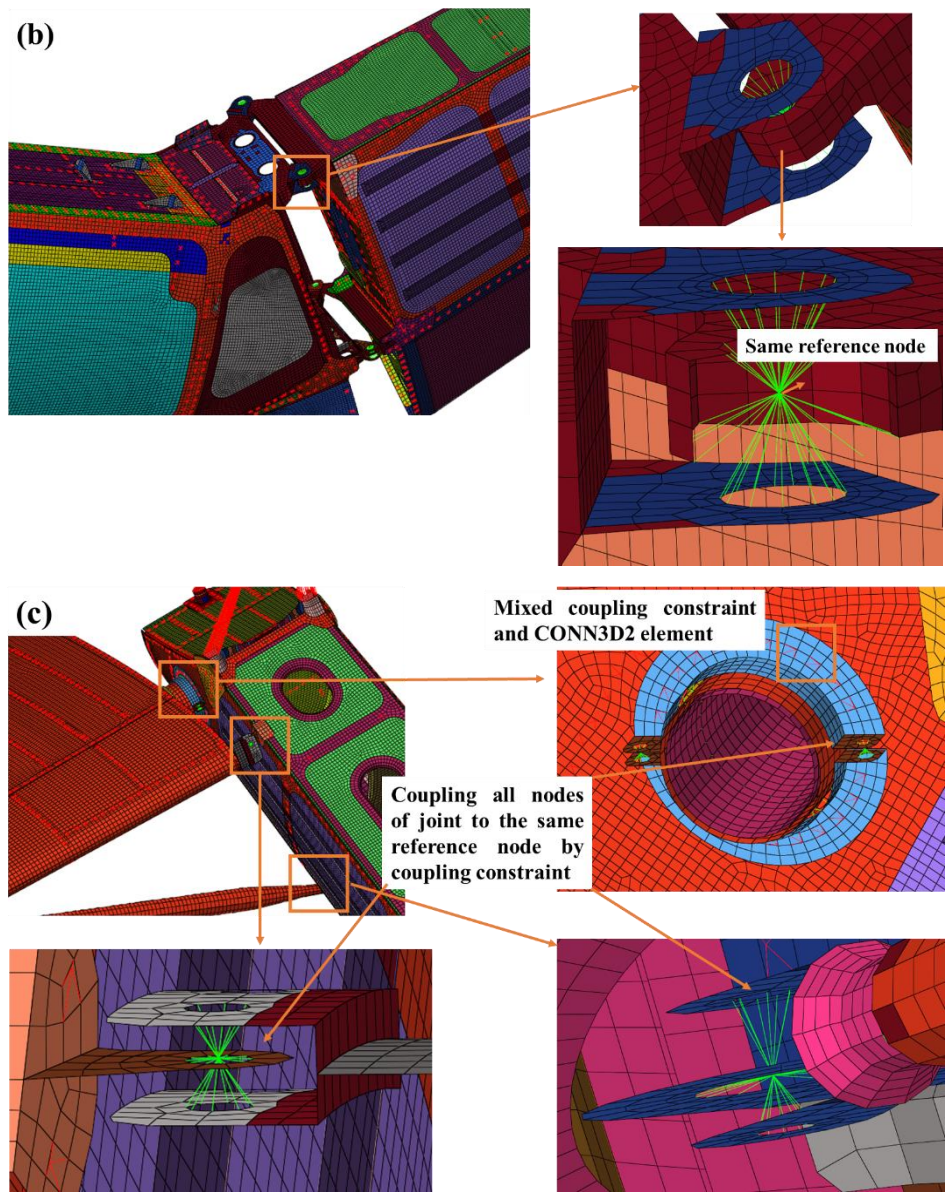

**Fig. A1** Connection strategies: (a) Connection between fuselage transition segment and tail segment; (b) Connection between tail segment and inclined beam; (c) Connection between inclined beam and horizontal tail.

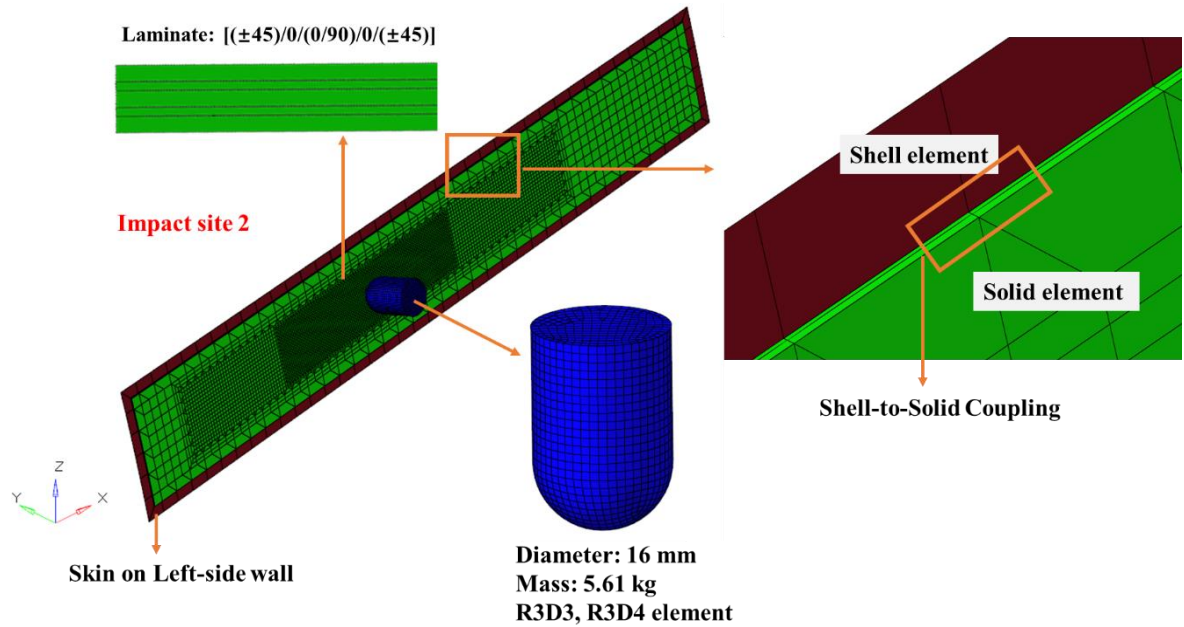

Fig. A2  Local FE model at impact site 2.

# Appendix B

Solution-dependent state variables (SDVs) are explained as listed Table B1.

Table B1  Explanations of solution-dependent state variables

| SDVs | Explanations |
|---|---|
| SDV1 | warp fibre tension failure of plain-weave (PW) or fibre tension failure of unidirectional (UD) |
| SDV2 | warp fibre compression failure of PW or fibre compression failure of UD |
| SDV3 | weft fibre tension failure of PW or matrix tension failure of UD |
| SDV4 | weft fibre compression failure of PW or matrix compression failure of UD |
| SDV5 | tension delamination for PW or UD |
| SDV6 | compression delamination for PW or UD |
| SDV14 | warp fibre tension strength degradation of PW or fibre tension strength degradation of UD |
| SDV15 | warp fibre compression strength degradation of PW or fibre compression strength degradation of UD |
| SDV16 | weft fibre tension strength degradation of PW or matrix tension strength degradation of UD |
| SDV17 | weft fibre compression strength degradation of PW or matrix compression strength degradation of UD |
| SDV18 | through-thickness tension strength degradation of PW or UD |
| SDV19 | through-thickness compression strength degradation of PW or UD |
| SDV20 | in-plane shear strength degradation of PW or UD in longitudinal-transverse direction |
| SDV21 | out-plane shear strength degradation of PW or UD in transverse-through thickness direction |
| SDV22 | out-plane shear strength degradation of PW or UD in longitudinal-through thickness direction |